\documentclass[reprint,amsmath,amssymb,aps,prd,groupedaddress,showpacs,nofootinbib]{revtex4-1}

\usepackage{color}

\usepackage{bm}
\usepackage{graphicx}
\usepackage{dcolumn}
\newcommand{\be}{\begin{equation}}
\newcommand{\ee}{\end{equation}}
\newcommand{\bea}{\begin{eqnarray}}
\newcommand{\eea}{\end{eqnarray}}

\newcommand{\dbss}[1]{_{\scriptstyle #1}}
\newcommand{\linda}[1]{\vphantom{A^{B}} #1}

\newcommand{\lindc}[1]{\vphantom{A^{B^j}} #1}

\newcommand{\mbss}[1]{_{\mbox{\scriptsize #1}}}

\newcommand{\mbsu}[1]{\mbox{\scriptsize #1}}
\newcommand{\mbtu}[1]{\mbox{\tiny #1}}

\newcommand{\ve}{\varepsilon}

\newcommand{\vphi}{\varphi}

\newcommand{\vtld}{\tilde{v}}
\newcommand{\etld}{\tilde{e}}
\newcommand{\gtld}{\tilde{g}}

\newcommand{\vpb}{\vphantom{\beta}}

\newcommand{\vptl}{\vphantom{\lambda}}

\newcommand{\vpsq}{\vphantom{S^2}}

\newcommand{\vphu}{\vphantom{*}}
\newcommand{\mfe}{\mathfrak{E}}
\newcommand{\mft}{\mathfrak{T}}

\begin{document}

\title{Dual relativity and its cosmological consequences}

\author{V. I. Tselyaev}
\affiliation{St. Petersburg State University, St. Petersburg, 199034, Russia}
\email{tselyaev@mail.ru}
\date{\today}

\begin{abstract}
A new version of the modified theory of gravity is formulated
in which two physical metrics are constructed out of two vierbeins
connected with each other by the duality condition including the flat metric
of the prior geometry.
The duality condition plays a crucial role in this theoretical scheme, and thus
gives the name to the whole approach: the theory of dual relativity (TDR).
The energy-momentum tensor density of a closed system is defined,
and the conservation law for this tensor density is deduced.
In the TDR it is assumed that there exist two kinds of matter governed by the
mutually dual physical metrics, and it is shown that the interaction between them
has antigravitational character.
A cosmological limit
of the field equations of the TDR is considered.
In this limit, there are two types of solutions: with positive
and with negative energy density. The first type admits the existence of
a stable Universe as a whole. The second type gives the oscillations
of the cosmological scale factor within the finite limits excluding the zero point,
that can be treated as a solution for a certain domain of the Universe.
For this type of solutions, the model formula for the dependence of the Hubble parameter $H$
on the redshift $z$ has been obtained. The values of the parameters of this formula are found
from the fit to the available $H(z)$ data. It is obtained that the TDR gives
a better description of the $H(z)$ data as compared to the flat $\Lambda$CDM model.
The important consequences of the obtained results are, first, their incompatibility
with the standard Big Bang model and, second, the existence of two critical values
of the scale factor determining the points of its sharp change in the course of oscillations
in the solutions of the second type.
\end{abstract}


\maketitle

\section{Introduction}

A unified description of the gravitational effects and/or the effects related to
the space-time geometry in a wide range of scales
is an important and long-standing problem of the theoretical physics.
Within the classical general theory of relativity \cite{Einstein16}
(GR in the conventional terminology)
this problem is successfully solved on the macroscopic scales
(of order the scale of the solar system)
where the GR ``has managed to survive every test'' \cite{Will14} without introducing
any fitting parameters.
A large number of models were developed on the basis of the GR to describe
data of the astrophysical observations on the cosmological scales
(see Refs.~\cite{Clifton12,Joyce15,Heisenberg19} for the review).
But here the classical GR equations need modifications which in the simplest case
reduce to adding the so-called $\Lambda$-term that represents the contribution
of the {\it dark energy} \cite{Overduin04} into the gravitational action.
In particular, the $\Lambda$CDM model ($\Lambda$-term plus cold dark matter,
\cite{Peebles84}) is frequently used in the analysis of $H(z)$ data,
i.e., in the description of the measured dependence
of the Hubble parameter $H$ on the redshift $z$.
However, the models which include $\Lambda$-term face the known problem
(see Ref.~\cite{Martin12} for its detailed discussion)
of the huge difference between the small ``observed'' value of the constant $\Lambda$
in the cosmology and the large estimated value of the contribution of the vacuum energy
of the matter fields into this quantity.

Besides the practical problems of describing the data,
there are the important questions connected with the internal consistency of the theory
and possibilities of its extension. One of them is the question of the {\it prior geometry}.
It is assumed that this geometry is determined by the metric tensor which is introduced
in the theory along with
the tensor $g$ of the GR, but which, in contrast with $g$, is not a dynamical variable.
The introduction of such additional (background) metric is at variance with ideology of the
classical GR, and for this reason, and also in view of the nice agreement of the GR with data,
it is sometimes an object of criticism, see, e.g., Ref.~\cite{Misner73}.
But on the one hand, the discrepancy (if any) between the results of the modified theories and
the predictions of the GR on the macroscopic scales
depends largely on the way in which this background metric is introduced.
On the other hand, there are substantial reasons to include the non-dynamical metric in the theory.

First, the energy-momentum tensor $T$ of any physical system in the GR
is defined via the functional derivative of its action functional $S$ with respect to
the metric tensor $g$, i.e., $T \sim \delta S/ \delta g$, where $g$ is treated as an external field.
If this definition is formally applied to the system ``matter plus gravitational field'',
one obtains that for this system $T$ is exactly equal to zero by virtue of the equations of motion.
Since this result is obviously unsatisfactory from the physical point of view
(see Ref.~\cite{Logunov89} where this issue is discussed in detail),
the so-called energy-momentum pseudotensor for the gravitational field is usually introduced,
but its definition is ambiguous and its specific properties are expressed by the prefix
``pseudo'', see \cite{Misner73,Landau71}.
This problem is solved in some modern modifications of the GR,
see Refs. \cite{Jimenez19,Heisenberg24},
and, as was shown, e.g., in Refs. \cite{Logunov89,Rosen85}, can be in principle
solved in the models including the background metric.

Second, the classical GR is the theory in which the carrier of the gravitational force,
graviton, is massless. A natural extension of the GR is the class of theories
with massive graviton. In most cases these theories,
which have a long history (see Refs.~\cite{Joyce15,Hinterbichler} for the review),
include an additional non-dynamical metric as indispensable element.
Notice that generally they can lead to the problems of discontinuity
\cite{vanDam70,Zakharov70,Vainshtein72} (see also Ref.~\cite{Babichev10}
where this problem was investigated numerically) and ghosts \cite{Boulware72}.
A consistent theory of massive gravity which is free of these drawbacks
was constructed in Refs. \cite{deRham10,deRham11,Hassan12}.

The other interesting questions are related to the properties of solutions of the equations
of modified gravity theories in the cosmological limit.
This refers to the possibility of the existence of such solutions
which correspond to the static, cyclic, or oscillating Universe
\cite{Goswami08,Parisi12,Bars11,Ijjas22,Maeda10,Zhang13}.
In particular, the oscillating solutions
\cite{Maeda10,Zhang13} are interesting from the point of view of possibility
to avoid a Big Bang initial singularity problem.
However, the correspondence of these solutions with data of the observations
is an open issue.

The aim of the present paper is an attempt to construct a new version of the
modified gravity theory with the background metric,
in which some of the above-mentioned problems are solved or, possibly, can be solved,
and which is compatible with data of the astrophysical observations
on the cosmological scales.
The special attention is paid to the definition and properties
of the total energy-momentum tensor density of the closed physical system.
The cosmological limit of the model equations is considered in which the static and
oscillating solutions are discussed. The latter class of the solutions is applied
to the analysis of the available $H(z)$ data.
The paper is organized as follows.
In Sec.~\ref{sec:basdef}, the basic definitions and some preliminary comments are given.
In Sec.~\ref{sec:act}, the total action functional of the theory is determined.
The equations of motion obtained from this functional,
definition of the total energy-momentum tensor density, and the respective
conservation laws are presented and discussed in Sec.~\ref{sec:eom}.
Section~\ref{sec:coslim} contains the equations of the theory in the cosmological limit,
their analysis, and the numerical results obtained from the fit of the model
parameters to the data of the observations.
Summary and conclusions are given in the last section.

\section{Basic definitions and preliminaries
\label{sec:basdef}}

The physical space-time is supposed to be a 4D pseudo-Riemannian
manifold ${\cal{M}}_{\linda{4}}$. The vectors and tensors on this manifold
and its coordinates $x^{\mu}$ will be labelled by the Greek letters
$\mu,\nu,\ldots \in \{1,2,3,4\}$ (holonomic indices).
The background metric $\gamma^{\vphu}_{\mu\nu}$ on ${\cal{M}}_{\linda{4}}$
is flat and has the signature $-2$.
The lower case Latin letters $\,a, b, \ldots \in \{1,2,3,4\}$
indicate anholonomic indices of the local frames being the vectors
in the flat 4D Minkowski space $M(3,1)$ with metric $\eta^{\vphu}_{ab}$,
\be
\eta^{\vphu}_{ab} = \mbox{diag}\,\{-1,-1,-1,+1\}\,.
\label{defeta}
\ee
The metric $\eta^{\vphu}_{ab}$ is used in what follows for
manipulations involving these 4D indices.
The symbols $\ve^{\kappa\lambda\mu\nu}$ and $\ve^{\vphu}_{abcd}$
stand for the totally antisymmetric tensors on ${\cal{M}}_{\linda{4}}$
and $M(3,1)$, respectively, with
$\ve^{1234} = -1$ and $\ve^{\vphu}_{1234} = 1$.

The main dynamical variables of the theory are two vierbein (or local frame) fields
$e^a_{\mu}$ and $\etld^a_{\mu}$ and the Lorentz spin-connection field
$\omega^{ab}_{\mu} = -\omega^{ba}_{\mu}$, being
the gauge field of the local Lorentz $SO(3,1)$ transformations.
The vierbeins $e^a_{\mu}$ and $\etld^a_{\mu}$ determine two metrics
$g_{\dbss{\mu\nu}}$ and $\gtld_{\dbss{\mu\nu}}$ according to the usual relations
\be
g_{\dbss{\mu\nu}} = e^a_{\mu} e_{\dbss{a\nu}},
\quad
\gtld_{\dbss{\mu\nu}} = \etld^a_{\mu} \etld_{\dbss{a\nu}}.
\label{ggtstg}
\ee
It is assumed that the vierbein fields satisfy the equation
\be
e^a_{\mu}\,\etld_{\dbss{a\nu}} = \gamma_{\dbss{\mu\nu}}\,,
\label{dualc}
\ee
that will be called the duality condition.
This condition leads to the following
relations between the metrics $g_{\dbss{\mu\nu}}$ and
$\gtld_{\dbss{\mu\nu}}$
\be
\gtld_{\dbss{\mu\nu}} =
\gamma_{\dbss{\mu\lambda}} g^{\lambda\kappa} \gamma_{\dbss{\kappa\nu}}\,,
\qquad
g_{\dbss{\mu\nu}} =
\gamma_{\dbss{\mu\lambda}} \gtld^{\lambda\kappa} \gamma_{\dbss{\kappa\nu}}\,,
\label{ggtrel}
\ee
where the inverse matrices $g^{\mu\nu}$, $\gtld^{\mu\nu}$, and $\gamma^{\mu\nu}$
are determined by the equations
\be
g^{\mu\lambda} g_{\dbss{\lambda\nu}} =
\gtld^{\mu\lambda} \gtld_{\dbss{\lambda\nu}} =
\gamma^{\mu\lambda} \gamma_{\dbss{\lambda\nu}} = \delta^{\mu}_{\dbss{\nu}}
\label{definvg}
\ee
and the invertibility of the $4 \times 4$ matrices $e^a_{\mu}$ and $\etld^a_{\mu}$
is supposed.
Since the given scheme includes several metric tensors,
the operations of the raising and lowering of the holonomic indices
are not unambiguously determined. So, the respective definitions will be given
in every case where this is necessary.

The theory developed in the following sections contains the additional degrees of freedom
as compared to the GR due to introducing two vierbein fields.
However, since the background metric $\gamma^{\vphu}_{\mu\nu}$ is fixed
(up to the proper coordinate transformation),
the condition (\ref{dualc}) reduces the number of the independent field variables
of the theory to the number of variables of the GR.
Nevertheless, the question remains concerning the grounds to include the additional
vierbein field from the very beginning.
Below the reasons of the origin of the second vierbein are explained.

In Ref.~\cite{T08} it was shown that it is possible to construct the action
functional of the 4D theory which is invariant under the
general linear non-degenerated transformations of four-component Majorana spinors
forming the group denoted in~\cite{T08} by $GL(4,M)$.
It was also shown that:
(i) $GL(4,M)$ is isomorphic to $GL(4,R)$ and includes $SL(2,C)$;
(ii) $GL(4,M)$ is the covering of the group of linear transformations of the vectors
and covectors
in the 6D Minkowski space $M(3,3)$ including the so-called {\it superluminal} transformations;
(iii) the spinor part of the action of the extended theory reduces to the spinor action
of the standard $SL(2,C)$-invariant 4D theory under a certain choice of the gauge.
However, to construct the spinor action of the extended theory it is necessary
to introduce two extended local frames (or vielbeins) of the form $e^K_{\mu}$ and
$\etld_{\dbss{K\mu}}$ (with $K, L, \ldots \in \{1,2,3,4,5,6\}$)
which are the vectors and covectors in $M(3,3)$.

There are various ways to introduce these extended frames, one of which was
considered in Ref.~\cite{T08}.
In this approach, the first vielbein was treated as the independent variable,
while the second was expressed through the first by means of the $SO(3,3)$-scalar field.
There is, however, a class of theories in which the vielbeins are deduced
from the gauge fields of some symmetry group with the help of the auxiliary fields-artifacts
(see, e.g., Refs. \cite{Stelle80,Pagels84,Kerrick95} and references therein).
In the case under consideration, this treatment suggests that
the extended frames are generated by
the gauge fields of the group $GL(4,M)$, while the role of the auxiliary fields is played
by two fundamental fields: vector $v^K$ and covector $\vtld^{\vphu}_K$
in $M(3,3)$ which are supposed to be dependent on the point of
${\cal{M}}_{\linda{4}}$.
Then, covariant derivatives of these fields $v^K_{:\,\mu}$ and $\vtld_{\dbss{K:\,\mu}}$
determine the vielbeins as follows
\be
e^K_{\mu} = \xi v^K_{:\,\mu}\,,\qquad
\etld_{\dbss{K\mu}} = \tilde{\xi} \vtld_{\dbss{K:\,\mu}},
\label{defexiv}
\ee
where $\xi$ and $\tilde{\xi}$ are the scalar fields having dimension of length.
Duality condition (\ref{dualc}) is generalized in this case in a natural way as
\be
e^K_{\mu}\,\etld_{\dbss{K\nu}} = \gamma_{\dbss{\mu\nu}}\,.
\label{gdualc}
\ee
Conditions of the orthogonality and normalization imposed on the fields
$v^K$ and $\vtld^{\vphu}_K$ and the choice of the appropriate gauge
enable one to eliminate these fields from the given scheme,
after which Eqs. (\ref{defexiv}) take the following form
\begin{subequations}
\bea
e^a_{\mu} &=& \xi \Omega^{6a}_{\mu}\,,\qquad
e^5_{\mu} = e^6_{\mu} = 0\,,
\label{deffrm1}\\
\etld^a_{\mu} &=& \tilde{\xi} \Omega^{5a}_{\mu}\,,\qquad
\etld^5_{\mu} = \etld^6_{\mu} = 0\,,
\label{deffrm2}
\eea
\label{deffrm0}
\end{subequations}
where $\Omega^{KL}_{\mu}$ are the components of the gauge field
(spin connection) of the group $EL(3,3)$ defined in Ref.~\cite{T08}.
Thus, Eqs. (\ref{defexiv}) and (\ref{deffrm0}) elucidate the origin
of two local frames in the extended theory and their reduction to the veirbeins
in $M(3,1)$ space.
However, consistent treatment of this approach is beyond the scope of the present paper.

\section{Action functional in dual relativity
\label{sec:act}}

The total action functional of the theory which will be called in the following
the theory of dual relativity (TDR) is taken in the form
\be
S^{\vphu}_{\mbsu{tot}} = S^{\vphu}_{\mbsu{g}} + S^{\vphu}_{\mbsu{m}}
+ \tilde{S}^{\vphu}_{\mbsu{m}}+ S^{\vphu}_{\mbsu{c}}\,,
\label{acttot}
\ee
where $S^{\vphu}_{\mbsu{g}}$ is the gravitational action,
$S^{\vphu}_{\mbsu{m}}$ is the action of the fields of ordinary matter
(governed by the metric $g_{\dbss{\mu\nu}}$),
$\tilde{S}^{\vphu}_{\mbsu{m}}$ is the action of the fields of dual matter
(governed by the metric $\gtld_{\dbss{\mu\nu}}$), and
$S^{\vphu}_{\mbsu{c}}$ is the term introduced to ensure fulfillment
of the duality condition (\ref{dualc}).

The gravitational action consists of the kinetic and interaction terms:
\be
S^{\vphu}_{\mbsu{g}} = S^{\,\mbsu{kin}}_{\mbsu{g}} + S^{\,\mbsu{int}}_{\mbsu{g}}\,.
\label{actg}
\ee
The kinetic term is taken in the form of the action of the Utiyama-Kibble
gauge theory of gravity \cite{Utiyama56,Kibble61}
(known also as Hilbert-Palatini action)
in which the vierbein fields and the spin-connection field are the
independent variables. It reads
\bea
\hspace{-3em}
S^{\,\mbsu{kin}}_{\mbsu{g}} &=& S^{\,\mbsu{kin}}_{\mbsu{g}}[\,e,\etld,\omega\,]
\nonumber\\
&=& -\frac{1}{8 c \kappa_{\mbsu{g}}} \int d^{4}x\,
\ve^{\kappa\lambda\mu\nu} \ve^{\vphu}_{abcd}\,
e^{(\zeta)a}_{\vptl\kappa} e^{(\zeta)b}_{\lambda} \omega^{cd}_{\vptl\mu\nu}\,,
\label{actgk}
\eea
where
\be
\kappa_{\mbsu{g}} = \frac{8\pi G_{\mbsu{N}}}{c^4} = \frac{8\pi\hbar}{m^2_{\mbsu{P}}c^3}
= \frac{8\pi l^2_{\mbsu{P}}}{\hbar c}\,,
\label{def:kg}
\ee
$G_{\mbsu{N}}$ is Newtonian constant, $m^{\vphu}_{\mbsu{P}}$ is the Planck mass,
$l^{\vphu}_{\mbsu{P}}$ is the Planck length,
\be
e^{(\zeta)a}_{\mu} = e^a_{\mu} + \zeta\,\etld^a_{\mu}\,,
\label{dez}
\ee
$\zeta$ is a mixing parameter,
$\omega^{ab}_{\mu\nu}$ is the strength tensor of the field $\omega^{ab}_{\mu}$:
\bea
\omega^{ab}_{\mu\nu} &=&
\partial_{\dbss{\mu}} \omega^{ab}_{\nu} - \partial_{\dbss{\nu}} \omega^{ab}_{\mu}
\nonumber\\
&+& \bigl(\,\omega^{aa'}_{\nu} \omega^{b'b}_{\mu} -
\omega^{aa'}_{\mu} \omega^{b'b}_{\nu}\,\bigr)\,\eta^{\vphu}_{a'b'}\,.
\label{domst}
\eea
The interaction term reads
\bea
\hspace{-3em}
S^{\,\mbsu{int}}_{\mbsu{g}} &=& S^{\,\mbsu{int}}_{\mbsu{g}}[\,e,\etld\,]
\nonumber\\
&=& \frac{1}{c \kappa_{\mbsu{g}}} \int d^{4}x\,
(\lambda_{31}\mfe_{31} + \lambda_{22}\mfe_{22} + \lambda_{13}\mfe_{13})
\label{actgi}
\eea
where
\begin{subequations}
\bea
\mfe_{31} &=& -\frac{1}{24} \ve^{\kappa\lambda\mu\nu} \ve^{\vphu}_{abcd}\,
e^a_{\vptl\kappa} e^b_{\lambda} e^c_{\vptl\mu} \etld^d_{\vptl\nu}\,,
\label{de31}\\
\mfe_{22} &=& -\frac{1}{24} \ve^{\kappa\lambda\mu\nu} \ve^{\vphu}_{abcd}\,
e^a_{\vptl\kappa} e^b_{\lambda} \etld^c_{\vptl\mu} \etld^d_{\vptl\nu}\,,
\label{de22}\\
\mfe_{13} &=& -\frac{1}{24} \ve^{\kappa\lambda\mu\nu} \ve^{\vphu}_{abcd}\,
e^a_{\vptl\kappa} \etld^b_{\lambda} \etld^c_{\vptl\mu} \etld^d_{\vptl\nu}\,,
\label{de13}
\eea
\label{def:e123}
\end{subequations}
$\lambda_{31}$, $\lambda_{22}$, and $\lambda_{13}$ are the constants.
The action $S^{\vphu}_{\mbsu{g}}$ is invariant under the local $SO(3,1)$
transformations and
is a polynomial in all the field variables and their first derivatives.

The actions of the matter fields have the form
\begin{subequations}
\bea
S^{\vphu}_{\mbsu{m}} &=& S^{\vphu}_{\mbsu{m}}[\,e,\vphi]
\nonumber\\
&=& \int d^{4}x\,\left( {\cal L}_{\,\mbsu{m}}
+ \frac{\lambda_{40}}{c \kappa_{\mbsu{g}}}\,\mfe_{40} \right),
\label{actom}\\
\tilde{S}^{\vphu}_{\mbsu{m}} &=& \tilde{S}^{\vphu}_{\mbsu{m}}[\,\etld,\tilde{\vphi}]
\nonumber\\
&=& \int d^{4}x\,\left( \tilde{{\cal L}}_{\,\mbsu{m}}
+ \frac{\lambda_{04}}{c \kappa_{\mbsu{g}}}\,\mfe_{04} \right),
\label{actdm}
\eea
\label{def:actm}
\end{subequations}
where $\vphi$ and $\tilde{\vphi}$ represent the field variables
of the ordinary and dual matter, ${\cal L}_{\,\mbsu{m}}$ and $\tilde{{\cal L}}_{\,\mbsu{m}}$
stand for their Lagrangian densities,
\begin{subequations}
\bea
\mfe_{40} &=& -\frac{1}{24} \ve^{\kappa\lambda\mu\nu} \ve^{\vphu}_{abcd}\,
e^a_{\vptl\kappa} e^b_{\lambda} e^c_{\vptl\mu} e^d_{\vptl\nu}\,,
\label{de40}\\
\mfe_{04} &=& -\frac{1}{24} \ve^{\kappa\lambda\mu\nu} \ve^{\vphu}_{abcd}\,
\etld^a_{\vptl\kappa} \etld^b_{\lambda} \etld^c_{\vptl\mu} \etld^d_{\vptl\nu}\,,
\label{de04}
\eea
\label{de4004}
\end{subequations}
$\lambda_{40}$ and $\lambda_{04}$ are the constants.
It should be emphasized that the functional $S^{\vphu}_{\mbsu{m}}$
is supposed to be independent on the fields $\etld^a_{\mu}$ and $\omega^{ab}_{\mu}$,
while $\tilde{S}^{\vphu}_{\mbsu{m}}$
is supposed to be independent on the fields $e^a_{\mu}$ and $\omega^{ab}_{\mu}$.
The independence of the matter actions on the spin connection $\omega^{ab}_{\mu}$
requires clarification in the case of their spinor parts. Here it is supposed
that in these parts of the actions the spin connection is replaced by
the Ricci rotation coefficients which are expressed via the vierbeins and their
derivatives (see \cite{Misner73,Wald84}).

The quantities $\mfe_{31}$, $\mfe_{22}$, $\mfe_{13}$, $\mfe_{40}$, and $\mfe_{04}$ defined by
Eqs. (\ref{def:e123}) and (\ref{de4004}) are
the {\it pseudoscalar} densities, because their signs depend on the mutual orientation
of the bases in ${\cal{M}}_{\linda{4}}$ and $M(3,1)$.
This orientation can be chosen in such a way that the equalities
\be
\mfe_{40} = \sqrt{-\det (g)},\qquad \mfe_{04} = \sqrt{-\det (\gtld)}
\label{detg}
\ee
are fulfilled.
The terms proportional to $\mfe_{40}$ and $\mfe_{04}$ are included in the matter actions
(\ref{def:actm}) to compensate contributions of the
vacuum expectation values of the matter fields. It is supposed that the values of
the constants $\lambda_{40}$ (which formally corresponds to the cosmological
constant $\Lambda$ in the GR)
and $\lambda_{04}$ are such that these contributions
are fully annihilated in (\ref{def:actm}).
It solves, at least at the classical level, the cosmological constant problem
mentioned in the Introduction, because in the TDR the effective value of this constant
is exactly equal to zero.

The last term of Eq. (\ref{acttot}) is defined by the formula
\bea
S^{\vphu}_{\mbsu{c}} &=& S^{\vphu}_{\mbsu{c}}[\,e,\etld,\gamma,\mft\,]
\nonumber\\
&=& \frac{1}{2 c} \int d^{4}x\,
(e^a_{\mu}\,\etld_{\dbss{a\nu}} - \gamma_{\dbss{\mu\nu}})\,\mft^{\mu\nu},
\label{actdc}
\eea
where $\mft^{\mu\nu}$ is the Lagrange multiplier being the tensor density and
having dimension of the energy density. Here the following remark is necessary.
To maintain the inversion symmetry of the total action, it is assumed
that the quantity $\mft^{\mu\nu}$ possesses the inversion properties which are
analogous to the properties
of the pseudoscalar densities in Eqs. (\ref{def:e123}) and (\ref{de4004}).
So, to be more precise, $\mft^{\mu\nu}$ should be called the {\it pseudotensor} density.
But to simplify the terminology and to avoid confusion with
the term ``energy-momentum pseudotensor for the gravitational field''
mentioned in the Introduction, the prefix ``pseudo'' in the name of $\mft^{\mu\nu}$
will be omitted.

The constants $\lambda_{31}$, $\lambda_{22}$, and $\lambda_{13}$ in Eq. (\ref{actgi})
are constrained by two conditions:
\be
\delta S^{\,\mbsu{int}}_{\mbsu{g}}/ \delta e^a_{\vptl\mu} = 0\quad \mbox{and} \quad
\delta S^{\,\mbsu{int}}_{\mbsu{g}}/ \delta \etld^a_{\vptl\mu} = 0
\label{sintcond}
\ee
at
\be
g_{\dbss{\mu\nu}} = \,\gtld_{\dbss{\mu\nu}} = \,\gamma_{\dbss{\mu\nu}}\,,
\label{eqmet}
\ee
that makes $\gamma_{\dbss{\mu\nu}}$ an equilibrium metric.
The conditions (\ref{sintcond}) and (\ref{eqmet}) are satisfied if
the constants $\lambda_{31}$, $\lambda_{22}$, and $\lambda_{13}$
are related by the equations
\be
\lambda_{31} = \lambda_{13} = 12 \lambda_{\mbsu{g}},\qquad
\lambda_{22} = -24 \lambda_{\mbsu{g}},
\label{lamg}
\ee
where $\lambda_{\mbsu{g}}$ is a new constant.
It is easy to see that the action $S^{\,\mbsu{int}}_{\mbsu{g}}$ vanishes
in the equilibrium point (\ref{eqmet}) if Eqs. (\ref{lamg}) are valid.

\section{Equations of motion
\label{sec:eom}}

\subsection{General form of the equations of motion,\\
solution for the spin connection,\\
and graviton mass
\label{ssec:mg}}

Equations of motion for the fields $\omega^{ab}_{\mu}$, $e^a_{\mu}$, and $\etld^a_{\mu}$
are obtained from the following stationarity conditions
\bea
\delta S^{\vphu}_{\mbsu{tot}}/ \delta \omega^{ab}_{\mu} &=& 0\,,
\label{omgstat}\\
\delta S^{\vphu}_{\mbsu{tot}}/ \delta e^a_{\mu} &=& 0\,,
\label{estat}\\
\delta S^{\vphu}_{\mbsu{tot}}/ \delta \etld^a_{\mu} &=& 0\,.
\label{etstat}
\eea
The solution of Eq. (\ref{omgstat}) is
\bea
\omega^{ab}_{\mu} &=& \frac{1}{2}\,\Bigl( e^{(\zeta)a\nu} e^{(\zeta)b}_{[\mu,\nu]}
- e^{(\zeta)b\nu} e^{(\zeta)a}_{[\mu,\nu]}
\nonumber\\
&+& e^{(\zeta)}_{c\mu}\,e^{(\zeta)a\nu}\,e^{(\zeta)b\lambda}\,
e^{(\zeta)c}_{[\lambda,\nu]}\,\Bigr),
\label{omgsol}
\eea
where
\be
e^{(\zeta)a}_{[\mu,\nu]} = \partial_{\dbss{\nu}} e^{(\zeta)a}_{\mu} -
\partial_{\dbss{\mu}} e^{(\zeta)a}_{\nu}\,,
\label{def:ezanm}
\ee
\be
e^{(\zeta)a\mu} = g^{(\zeta)\mu\nu} e^{(\zeta)a}_{\nu},
\label{def:ezam}
\ee
\be
g^{(\zeta)}_{\mu\nu} = \,e^{(\zeta)a}_{\mu}\,e^{(\zeta)}_{a\nu}\,
= \,g_{\dbss{\mu\nu}} + 2\zeta \gamma_{\dbss{\mu\nu}}
+ \zeta^2 \gtld_{\dbss{\mu\nu}}\,,
\label{def:gzet}
\ee
\be
g^{(\zeta)}_{\mu\lambda}g^{(\zeta)\lambda\nu} = \delta^{\nu}_{\mu}\,.
\label{def:gzetinv}
\ee

Here it is important to note the following.
If one substitutes the solution (\ref{omgsol}) into the functional (\ref{actgk})
and then expresses the vierbein $\etld^a_{\vptl\mu}$ via $e^a_{\vptl\mu}$
with the use of Eqs. (\ref{ggtstg}), (\ref{dualc}), and (\ref{definvg}),
one obtains the functional $S^{\vphu}_{\mbsu{g}}$ that depends only on the metrics
$g_{\dbss{\mu\nu}}$ and $\gamma_{\dbss{\mu\nu}}$ and their derivatives.
In this case the kinetic part of $S^{\vphu}_{\mbsu{g}}$ reduces to the conventional
gravitational action of GR if one also sets $\zeta = 0$.
These transformations enable one to expand $S^{\vphu}_{\mbsu{g}}$ (at any $\zeta$)
in powers of the difference
$h_{\dbss{\mu\nu}} = g_{\dbss{\mu\nu}} - \gamma_{\dbss{\mu\nu}}$
that determines deviation of $g_{\dbss{\mu\nu}}$ from the equilibrium metric.
Using this expansion and assuming that Eqs. (\ref{lamg}) are fulfilled,
one can obtain that up to the second order terms in $h_{\dbss{\mu\nu}}$,
the action $S^{\vphu}_{\mbsu{g}}$ reduces to the Fierz-Pauli action
(see, e.g., Ref.~\cite{Hinterbichler}) with the graviton mass $m^{(\zeta)}_{\mbsu{g}}$
determined by the equations
\be
m^{(\zeta)}_{\mbsu{g}} = m^{(0)}_{\mbsu{g}}/\,(1-\zeta)\,,\qquad
m^{(0)}_{\mbsu{g}}c/\hbar = \sqrt{8 \lambda_{\mbsu{g}}}\,,
\label{gravmass}
\ee
where $m^{(0)}_{\mbsu{g}}$ denotes the graviton mass at $\zeta = 0$.
Obviously, Eqs. (\ref{gravmass}) imply that the constant $\lambda_{\mbsu{g}}$
should be non-negative.
Conversely, this constant can be expressed in terms of the Compton wavelength
of the graviton $l^{(0)}_{\mbsu{g}}$ as
\be
\lambda_{\mbsu{g}} = l^{(0)-2}_{\mbsu{g}}/8\,,\qquad
l^{(0)}_{\mbsu{g}} = \frac{\hbar}{m^{(0)}_{\mbsu{g}}\!c}\,.
\label{gravcwl}
\ee
Thus, the TDR is the theory with massive graviton.
The problems which can emerge in the theories of this type were mentioned
in the Introduction. However, in the present paper they are not analyzed.

\subsection{Equations of motion in asymmetric form
\label{ssec:eomas}}

Equations of motion (\ref{estat}) and (\ref{etstat}) contain unnecessary
degrees of freedom related to the Lorentz rotations and represented by
the anholonomic indices. One can remove these degrees of freedom by
replacing the anholonomic indices with holonomic ones.
It can be done with the help of the following equivalent transformation
of the system (\ref{estat}) and (\ref{etstat}):
\bea
&&e^a_{\nu}\,\delta S^{\vphu}_{\mbsu{tot}}/ \delta e^a_{\mu}
+ \etld^a_{\nu}\,\delta S^{\vphu}_{\mbsu{tot}}/ \delta \etld^a_{\mu} = 0\,,
\label{aseom1}\\
&&\etld^a_{\nu}\,\delta S^{\vphu}_{\mbsu{tot}}/ \delta \etld^a_{\mu} =
\zeta\,\etld^a_{\nu}\,\delta S^{\vphu}_{\mbsu{tot}}/ \delta e^a_{\mu}\,.
\label{aseom2}
\eea
After some algebra, these equations take the form:
\begin{subequations}
\bea
&& G^{(\zeta)\mu}_{\,\nu} = Y^{\mu}_{\nu} + Z^{\mu}_{\nu}\,,
\label{eomas1}\\
&& Z^{\mu}_{\nu} = \zeta\,Y^{\mu}_{\lambda} f^{\lambda}_{\nu}\,,
\label{eomas2}
\eea
\label{eomas}
\end{subequations}
where
\be
G^{(\zeta)\mu}_{\,\nu} =
R^{(\zeta)\mu}_{\,\nu} - \frac{1}{2}\,\delta^{\mu}_{\nu}\,R^{(\zeta)},
\label{def:gzmc}
\ee
$R^{(\zeta)\mu}_{\,\nu}$
is the conventional Ricci tensor in mixed components corresponding to the metric
$g^{(\zeta)}_{\mu\nu}$ and determined here by the formula
\be
R^{(\zeta)\mu}_{\,\nu} =
e^{(\zeta)\mu}_{\vptl a} e^{(\zeta)\lambda}_{\vptl b} \omega^{ab}_{\lambda\nu}
\label{rtmix}
\ee
[with $e^{(\zeta)\mu}_{\vptl a}$ and
$\omega^{ab}_{\mu\nu}$ given by Eqs. (\ref{domst}), (\ref{omgsol})--(\ref{def:gzetinv})],
$R^{(\zeta)}$ is the Ricci scalar ($R^{(\zeta)}=R^{(\zeta)\lambda}_{\,\lambda}$),
\begin{subequations}
\bea
\mfe_{\zeta}\,Y^{\mu}_{\nu} &=& \kappa_{\mbsu{g}}\Bigl(
\mfe_{40}\,T^{\mu}_{\nu} - \frac{1}{2} \mft^{\mu\lambda}\gamma_{\dbss{\lambda\nu}}\Bigr)
\nonumber\\
&+& \lambda_{\mbsu{g}}\Bigl( \mfe_{40}\,V^{\mu}_{\nu} + \mfe_{04}\,\tilde{W}^{\mu}_{\nu}\Bigr),
\label{def:ymn}\\
\mfe_{\zeta}\,Z^{\mu}_{\nu} &=& \kappa_{\mbsu{g}}\Bigl(
\mfe_{04}\,\tilde{T}^{\mu}_{\nu} - \frac{1}{2} \gamma_{\dbss{\nu\lambda}}\mft^{\lambda\mu}\Bigr)
\nonumber\\
&+& \lambda_{\mbsu{g}}\Bigl( \mfe_{04}\,\tilde{V}^{\mu}_{\nu} + \mfe_{40}\,W^{\mu}_{\nu} \Bigr),
\label{def:zmn}
\eea
\end{subequations}
\begin{subequations}
\bea
V^{\mu}_{\nu} &=& 3\,\bigl(\,f^{\mu}_{\nu} - f^{\lambda}_{\lambda}\,\delta^{\mu}_{\nu}\,\bigr)\,,
\label{def:vmn}\\
\tilde{V}^{\mu}_{\nu} &=& 3\,\bigl(\,\tilde{f}^{\mu}_{\nu}
- \tilde{f}^{\lambda}_{\lambda}\,\delta^{\mu}_{\nu}\,\bigr)\,,
\label{def:tvmn}
\eea
\end{subequations}
\begin{subequations}
\bea
W^{\mu}_{\nu} &=& 4\,\bigl(\,f^{\lambda}_{\lambda}\,f^{\mu}_{\nu}
- f^{\mu}_{\lambda}\,f^{\lambda}_{\nu}\,\bigr) -3\,f^{\mu}_{\nu}\,,
\label{def:wmn}\\
\tilde{W}^{\mu}_{\nu} &=& 4\,\bigl(\,\tilde{f}^{\lambda}_{\lambda}\,\tilde{f}^{\mu}_{\nu}
- \tilde{f}^{\mu}_{\lambda}\,\tilde{f}^{\lambda}_{\nu}\,\bigr) -3\,\tilde{f}^{\mu}_{\nu}\,,
\label{def:twmn}
\eea
\end{subequations}
\begin{subequations}
\bea
f^{\mu}_{\nu} &=& g^{\mu\lambda}\gamma_{\dbss{\lambda\nu}}
= \gamma^{\mu\lambda}\gtld_{\dbss{\lambda\nu}}\,,
\label{def:fmn}\\
\tilde{f}^{\mu}_{\nu} &=& \gamma^{\mu\lambda}g_{\dbss{\lambda\nu}}
= \gtld^{\mu\lambda}\gamma_{\dbss{\lambda\nu}}\,,
\label{def:tfmn}
\eea
\label{def:ftfmn}
\end{subequations}
\be
\mfe_{\zeta} = -\frac{1}{24}\,\ve^{\kappa\lambda\mu\nu} \ve^{\vphu}_{abcd}\,
e^{(\zeta)a}_{\vptl\kappa} e^{(\zeta)b}_{\lambda}
e^{(\zeta)c}_{\vptl\mu} e^{(\zeta)d}_{\vptl\nu},
\label{def:ezet}
\ee
$T^{\mu}_{\nu}$ and $\tilde{T}^{\mu}_{\nu}$ are the energy-momentum tensors
of the ordinary and dual matter, respectively, defined as
\begin{subequations}
\bea
\mfe_{40}\,T^{\mu}_{\nu} &=&
- c\,e^a_{\nu}\,\delta S^{\vphu}_{\mbsu{m}}/ \delta e^a_{\mu}\,,
\label{def:tmn}\\
\mfe_{04}\,\tilde{T}^{\mu}_{\nu} &=&
- c\,\etld^a_{\nu}\,\delta \tilde{S}^{\vphu}_{\mbsu{m}}/ \delta \etld^a_{\mu}\,.
\label{def:ttmn}
\eea
\label{def:tttmn}
\end{subequations}
Note also that the following relations are fulfilled
\begin{subequations}
\bea
\mfe_{40}\mfe_{04} &=& - \det (\gamma)\,,
\label{e4004rel}\\
\mfe_{\zeta}\mfe_{04} &=& - \det (\gamma + \zeta \gtld)\,,
\label{ez04rel}\\
\mfe_{\zeta}\mfe_{40} &=& - \det (g + \zeta \gamma)\,,
\label{ez40rel}
\eea
\label{ez4004rel}
\end{subequations}
that can be obtained from Eqs. (\ref{dualc}), (\ref{dez}), (\ref{de4004}),
and (\ref{def:ezet}).

\subsection{Symmetric form of the equations of motion
\label{ssec:eomsym}}

Equations (\ref{eomas}) have the asymmetric form
with respect to permutation of the holonomic indices.
This form is most appropriate for further analysis presented in Sec.~\ref{sec:coslim}
because it contains explicitly the Lagrange multiplier $\mft^{\mu\nu}$
for which the conservation law (see below Sec.~\ref{ssec:cl}) exists.
The symmetric form represents the TDR equations from another side and
is more convenient for comparison with the GR equations.
It can be obtained by eliminating $\mft^{\mu\nu}$ from Eqs. (\ref{eomas})
that results in the equation
\be
Q^{\kappa\lambda}_{\mu\nu}\,G^{(\zeta)}_{\kappa\lambda} = J^{\vphu}_{\mu\nu}\,,
\label{eomsym}
\ee
where
\be
G^{(\zeta)}_{\mu\nu} =
R^{(\zeta)}_{\mu\nu} - \frac{1}{2}\,g^{(\zeta)}_{\mu\nu}\,R^{(\zeta)}.
\label{def:gzet}
\ee
Here
$R^{(\zeta)}_{\mu\nu}$ is the symmetric Ricci tensor in covariant components
connected with tensor (\ref{rtmix}) by the formula
\be
R^{(\zeta)}_{\mu\nu} = g^{(\zeta)}_{\mu\mu'}\,R^{(\zeta)\mu'}_{\nu}.
\label{rtcov}
\ee
\be
Q^{\kappa\lambda}_{\mu\nu} =
q^{\kappa}_{\mu}\,(q^2)^{\lambda}_{\nu} +
(q^2)^{\kappa}_{\mu}\,q^{\lambda}_{\nu} -
q^{\kappa}_{\mu}\,q^{\lambda}_{\nu}\,,
\label{def:qbig}
\ee
the tensor $q^{\mu}_{\nu}$ is the solution of the equation
\be
\bigl( \delta^{\nu'}_{\nu} + \zeta\,f^{\nu'}_{\nu} \bigr)\,q^{\mu}_{\nu'} =
\delta^{\mu}_{\nu}\,,
\label{qfeq}
\ee
that is $q=(1+\zeta\,f)^{-1}$.
\bea
\mfe_{\zeta}\,J^{\vphu}_{\mu\nu} &=& \mfe_{40}\Bigl(\kappa_{\mbsu{g}}\,T^{\vphu}_{\mu\nu}
+ \lambda_{\mbsu{g}}\,U^{\vphu}_{\mu\nu}\Bigr)
\nonumber\\
&-& \mfe_{04}\,\tilde{f}^{\kappa}_{\mu}\,\tilde{f}^{\lambda}_{\nu}
\Bigl(\kappa_{\mbsu{g}}\,\tilde{T}^{\vphu}_{\kappa\lambda}
+ \lambda_{\mbsu{g}}\,\tilde{U}^{\vphu}_{\kappa\lambda}\Bigr),
\label{def:jmn}
\eea
\begin{subequations}
\bea
U^{\vphu}_{\mu\nu} &=& 6 \gamma_{\dbss{\mu\nu}} + 4 \gtld_{\dbss{\mu\nu}}
- f^{\lambda}_{\lambda}\,(4 \gamma_{\dbss{\mu\nu}} + 3 g_{\dbss{\mu\nu}})\,,
\label{def:umn}\\
\tilde{U}^{\vphu}_{\mu\nu} &=& 6 \gamma_{\dbss{\mu\nu}} + 4 g_{\dbss{\mu\nu}}
- \tilde{f}^{\lambda}_{\lambda}\,(4 \gamma_{\dbss{\mu\nu}} + 3 \gtld_{\dbss{\mu\nu}})\,,
\label{def:tumn}
\eea
\end{subequations}
\be
T^{\vphu}_{\mu\nu} = g_{\dbss{\mu\lambda}}\,T^{\lambda}_{\nu}\,,\qquad
\tilde{T}^{\vphu}_{\mu\nu} = \gtld_{\dbss{\mu\lambda}}\,\tilde{T}^{\lambda}_{\nu}\,,
\label{def:tmncov}
\ee
tensors $f^{\mu}_{\nu}$, $\tilde{f}^{\mu}_{\nu}$, $T^{\mu}_{\nu}$, and $\tilde{T}^{\mu}_{\nu}$
are defined in Eqs. (\ref{def:ftfmn}) and (\ref{def:tttmn}).

The TDR equation of motion (\ref{eomsym}) reduces to the GR equation
without the cosmological term if one sets the constants $\zeta$ and $\lambda_{\mbsu{g}}$
equal to zero and omits the energy-momentum tensor of the dual matter
$\tilde{T}^{\vphu}_{\mu\nu}$.
This means that to reproduce the GR results for gravitational effects
on the macroscopic scales,
the constants $\zeta$ and $\lambda_{\mbsu{g}}$ should be very small
(though in fact this may be insufficient because of the problems
arising in the theories with massive graviton, see \cite{Hinterbichler}).
But, as will be seen below, these constants should not be equal to zero exactly
and should be positive from the point of view of cosmological applications of the theory.

\subsection{Total energy-momentum tensor density\\
and the conservation laws
\label{ssec:cl}}

First of all, it should be noted that since the metric $\gamma_{\dbss{\mu\nu}}$
enters explicitly only into the term $S^{\vphu}_{\mbsu{c}}$ of the
total action functional, the following conservation law is fulfilled
\be
\partial_{\dbss{\mu}}\bigl(\,\mft^{(\mu\nu)}\gamma_{\dbss{\nu\lambda}}\,\bigr)
= \frac{1}{2}\,\mft^{(\mu\nu)}\,\partial_{\dbss{\lambda}}\gamma_{\dbss{\mu\nu}}\,,
\label{taucons}
\ee
where
\be
\mft^{(\mu\nu)} = \frac{1}{2}\,(\mft^{\mu\nu} + \mft^{\nu\mu})
\label{tausym1}
\ee
is the symmetric part of the Lagrange multiplier $\mft^{\mu\nu}$.
It can be proved by the known method (see, e.g., Ref.~\cite{Landau71})
using invariance of the total action under the
infinitesimal coordinate transformations.
However, in the given case a more simple way to prove (\ref{taucons})
is to use the fact that $\gamma_{\dbss{\mu\nu}}$ is the flat metric,
and therefore it can be represented in the form
\be
\gamma_{\dbss{\mu\nu}}(x) = \tilde{x}^a_{\mu}(x)\,\tilde{x}^b_{\nu}(x)\,\eta^{\vphu}_{ab}\,,
\label{gamrep}
\ee
where $\tilde{x}^a_{\mu}(x) = \partial_{\dbss{\mu}} \tilde{x}^a(x)$, and $\tilde{x}^a(x)$
are the arbitrary functions constrained only by the condition of invertibility
of the matrix $\tilde{x}^a_{\mu}$. The total action functional is invariant under
the variations of $\tilde{x}^a(x)$ because the metric (\ref{gamrep}) can be
always reduced to the diagonal form (\ref{defeta}) by means of the coordinate
transformation $x^{\mu} \rightarrow x^{\prime\mu}(x) = \delta^{\mu}_a\,\tilde{x}^a(x)$.
Then Eq. (\ref{taucons}) is readily obtained from the equality
\be
\delta S^{\vphu}_{\mbsu{tot}}/ \delta \tilde{x}^a = 0\,.
\label{xtcond}
\ee

According to Eq. (\ref{actdc}), the symmetric part of the Lagrange multiplier
$\mft^{\mu\nu}$ can be represented by the formula
\be
\mft^{(\mu\nu)} = -2c\,\delta S^{\vphu}_{\mbsu{tot}}/ \delta \gamma_{\dbss{\mu\nu}}\,,
\label{tausym2}
\ee
where variations of $\gamma_{\dbss{\mu\nu}}$ are not constrained by
the condition that this metric is flat.
Since $\gamma_{\dbss{\mu\nu}}$ is defined as the equilibrium metric,
it enables one to treat $\mft^{(\mu\nu)}$
as the total energy-momentum tensor density, and Eq.~(\ref{taucons})
as the law of its conservation in a closed system.

Though the value of $\mft^{(\mu\nu)}$ in each space-time point is unambiguously
determined by the equations of motion (see below),
the feasible explicit expressions for this quantity are not unique
because the Eq.~(\ref{eomsym}) [which does not contain $\mft^{(\mu\nu)}$]
can be used to get a multitude of the equivalent representations of this expression.
One of them can be obtained from Eq.~(\ref{eomas1}). It reads
\begin{subequations}
\bea
&& \hspace{-1em}
\mft^{(\mu\nu)} = \mft^{(\mu\nu)}_{\,\mbsu{m}} + \tilde{\mft}^{(\mu\nu)}_{\,\mbsu{m}}
+ \mft^{(\mu\nu)}_{\,\mbsu{g}}\,,
\label{tempd0}\\
&& \hspace{-1em}
\mft^{(\mu\nu)}_{\,\mbsu{m}} = \frac{1}{2}\mfe_{40}\bigl(\,
T^{\mu}_{\lambda}\gamma^{\lambda\nu}
+ T^{\nu}_{\lambda}\gamma^{\lambda\mu}\bigr),
\label{tempd1}\\
&& \hspace{-1em}
\tilde{\mft}^{(\mu\nu)}_{\,\mbsu{m}} = \frac{1}{2}\mfe_{04}\bigl(\,
\tilde{T}^{\mu}_{\lambda}\gamma^{\lambda\nu}
+ \tilde{T}^{\nu}_{\lambda}\gamma^{\lambda\mu} \bigr),
\label{tempd2}\\
&& \hspace{-1em}
\mft^{(\mu\nu)}_{\,\mbsu{g}} = -\frac{1}{2\kappa_{\mbsu{g}}}\bigl(
\mfe_{\zeta}\,G^{(\zeta)\mu}_{\,\lambda}\gamma^{\lambda\nu} +
\mfe_{\zeta}\,G^{(\zeta)\nu}_{\,\lambda}\gamma^{\lambda\mu} \bigr)
\nonumber\\
&& \hspace{-1em}
+ \frac{\lambda_{\mbsu{g}}}{2\kappa_{\mbsu{g}}}
\bigl[ \mfe_{40}\,\bigl( V^{\mu}_{\lambda} + W^{\mu}_{\lambda} \bigr)
+ \mfe_{04}\,\bigl( \tilde{V}^{\mu}_{\lambda} + \tilde{W}^{\mu}_{\lambda} \bigr) \bigr]
\gamma^{\lambda\nu}
\nonumber\\
&& \hspace{-1em}
+ \frac{\lambda_{\mbsu{g}}}{2\kappa_{\mbsu{g}}}
\bigl[ \mfe_{40}\,\bigl( V^{\nu}_{\lambda} + W^{\nu}_{\lambda} \bigr)
+ \mfe_{04}\,\bigl( \tilde{V}^{\nu}_{\lambda} + \tilde{W}^{\nu}_{\lambda} \bigr) \bigr]
\gamma^{\lambda\mu}.
\label{tempd3}
\eea
\label{tempd}
\end{subequations}
Here $\mft^{(\mu\nu)}_{\,\mbsu{m}}$, $\tilde{\mft}^{(\mu\nu)}_{\,\mbsu{m}}$,
and $\mft^{(\mu\nu)}_{\,\mbsu{g}}$
are the contributions of the ordinary and dual matter and of the gravity into the total
energy-momentum tensor density, respectively.

Nevertheless, in spite of the non-uniqueness of the general expression
for $\mft^{(\mu\nu)}$ mentioned above,
there is a unique representation of $\mft^{(\mu\nu)}$ that
does not contain derivatives of the metric tensors
(except for those which enter through the Ricci rotation coefficients
into the energy-momentum tensors of the matter).
It is obtained from Eq.~(\ref{eomas2}) and has the form of the following
linear algebraic equation for $\mft^{(\mu\nu)}$
\begin{subequations}
\be
\bigl(\delta^{\mu}_{\mu'} \delta^{\nu}_{\nu'}
- \zeta^2 f^{\mu}_{\mu'} f^{\nu}_{\nu'}\bigr)\,\mft^{(\mu'\nu')} =
\tilde{\mft}^{\mu\nu} + \tilde{\mft}^{\nu\mu},
\label{tempt0}
\ee
where
\bea
\tilde{\mft}^{\mu\nu} &=& \bigl(\delta^{\mu}_{\mu'} + \zeta f^{\mu}_{\mu'}\bigr)
\Bigl\{ \mfe_{04}\,\tilde{T}^{\mu'\nu'}\!f^{\nu}_{\nu'}
- \zeta\,\mfe_{40}\,T^{\mu'\nu}
\nonumber\\
&+& \frac{\lambda_{\mbsu{g}}}{\kappa_{\mbsu{g}}}
\Bigl[\,\bigl( \mfe_{40}\,W^{\mu'}_{\nu'} + \mfe_{04}\,\tilde{V}^{\mu'}_{\nu'} \bigr)
\gamma^{\nu'\nu}
\nonumber\\
&-& \zeta
\bigl( \mfe_{40}\,V^{\mu'}_{\nu'} + \mfe_{04}\,\tilde{W}^{\mu'}_{\nu'} \bigr)
g^{\nu'\nu}\Bigr] \Bigr\},
\label{tempt1}
\eea
\label{tempt}
\end{subequations}
and
\be
T^{\mu\nu} = T^{\mu}_{\lambda} g^{\lambda\nu},\qquad
\tilde{T}^{\mu\nu} = \tilde{T}^{\mu}_{\lambda} \gtld^{\lambda\nu}.
\label{def:tmncontv}
\ee
In the limit $\zeta \rightarrow +0$, one has from Eq. (\ref{tempt1})
\be
\tilde{\mft}^{\mu\nu} =
\mfe_{04}\,\tilde{T}^{\mu\nu'}\!f^{\nu}_{\nu'}
+ \frac{\lambda_{\mbsu{g}}}{\kappa_{\mbsu{g}}}
\Bigl( \mfe_{40}\,W^{\mu}_{\nu'} + \mfe_{04}\,\tilde{V}^{\mu}_{\nu'} \Bigr)
\gamma^{\nu'\nu}.
\label{tempt2}
\ee
If in addition the contribution of the dual matter is absent
($\tilde{T}^{\mu\nu} = 0$),
the total energy-momentum tensor density $\mft^{(\mu\nu)}$
in the representation (\ref{tempt})
is fully determined by the metric contribution,
which obviously should effectively include in this case
the contribution of the energy-momentum tensor of the ordinary matter.
This result illustrates the fact that the decomposition of $\mft^{(\mu\nu)}$
into the parts associated with matter (ordinary and dual) and gravity
[of the type used in Eqs. (\ref{tempd})] is relative.

Partial conservation laws for the matter can be formulated
if the sets of the field variables $\vphi$ and $\tilde{\vphi}$
of the ordinary and dual matter
in Eqs. (\ref{def:actm}) have no common elements
(i.e. if $\{\vphi\} \cap \{\tilde{\vphi}\} = \varnothing$).
In this case one has the following equalities for the tensors
$T^{\mu}_{\nu}$ and $\tilde{T}^{\mu}_{\nu}$ [cf. Eq.~(\ref{taucons})]
\begin{subequations}
\bea
\partial_{\dbss{\mu}}\bigl(\,\mfe_{40}\,T^{\mu}_{\lambda}\bigr)
&=& \frac{1}{2}\,\mfe_{40}\,T^{\mu\nu}
\partial_{\dbss{\lambda}}g_{\dbss{\mu\nu}}\,,
\label{tmncons}\\
\partial_{\dbss{\mu}}\bigl(\,\mfe_{04}\,\tilde{T}^{\mu}_{\lambda}\bigr)
&=& \frac{1}{2}\,\mfe_{04}\,\tilde{T}^{\mu\nu}
\partial_{\dbss{\lambda}}\gtld_{\dbss{\mu\nu}}\,,
\label{ttmncons}
\eea
\end{subequations}
which can be proved in the same way as in the conventional GR.
However,
it is worth noting that, in contrast to the GR equations,
Eqs. (\ref{eomas}) formally admit solution
with absent gravitational field in the presence of the matter.
This solution is given by Eqs. (\ref{eqmet}) and by the equalities
\be
\mft^{\mu\nu} = \mft^{\nu\mu} =
\mfe_{40}\,T^{\mu\nu} + \mfe_{04}\,\tilde{T}^{\mu\nu}
\label{eqsol1}
\ee
under the condition
\be
T^{\mu\nu} = \tilde{T}^{\mu\nu}.
\label{eqcond}
\ee
Equation (\ref{eqsol1}) also explains the choice of the overall factor
in Eq.~(\ref{actdc}) including the Lagrange multiplier $\mft^{\mu\nu}$.

\section{TDR equations in the cosmological limit
\label{sec:coslim}}

\subsection{Energy-momentum tensors of matter
\label{ssec:cl1}}

In the cosmological limit it is assumed that, first, the coordinate basis
in ${\cal{M}}_{\linda{4}}$ is chosen in such a way that the flat metric
$\gamma_{\dbss{\mu\nu}}$ has the diagonal form (\ref{defeta}) and, second,
the metric $g_{\dbss{\mu\nu}}$ in this basis satisfies the conditions of isotropy
and homogeneity, i.e. it has the form
of the flat Friedmann-Robertson-Walker (FRW) metric
\be
g_{\dbss{\mu\nu}} = \mbox{diag}\,\{-A^2,-A^2,-A^2,\,B^2\}\,,
\label{def:gmncl}
\ee
where the functions $A$ (scale factor) and $B$ (lapse) depend only
on the time-like coordinate $x^4$.

For the energy-momentum tensors of matter, the model of the perfect fluid
is used. In the case of the ordinary matter this tensor in mixed components
has a form
\be
T^{\mu}_{\nu} = (\ve + p)\,u^{\mu} u_{\dbss{\nu}} - p\,\delta^{\mu}_{\nu}\,,
\label{def:tpf}
\ee
where $\ve$ is the energy density, $p$ is the pressure, and $u^{\mu}$ is 4-velocity.
A comoving frame is chosen so that
\be
u^k = 0, \quad k=1,2,3\,;\qquad  u^{4} u_{\dbss{4}} = 1\,.
\label{ucomfr}
\ee
It is supposed that in the general case the perfect fluid consists of several species
in which $p$ and $\ve$ are related by the equation of state of the form
\be
p = \frac{\alpha}{3}\,\ve\,,
\label{peeom}
\ee
where $\alpha$ is a parameter that can take the values from 0 (dust-like matter)
to 1 (radiation). On the other hand,
the conservation law (\ref{tmncons}) also gives the relation between $p$ and $\ve$
which in the case of the metric (\ref{def:gmncl}) reads
\be
p(A) = -\ve(A) - \frac{A}{3}\,\frac{d\ve(A)}{dA}\,.
\label{perelat}
\ee
Substitution of Eq. (\ref{peeom}) into Eq. (\ref{perelat}) gives the following
$A$-dependence of $\ve$:
\be
\ve = \ve_{\dbss{\alpha}} = \frac{3\lambda_{\mbsu{g}}}{\kappa_{\mbsu{g}}}\,
\mu_{\dbss{\alpha}} A^{-\alpha-3},
\label{epsalp}
\ee
where $\mu_{\dbss{\alpha}}$ is a constant parameter and normalization multiplier
was chosen to make $\mu_{\dbss{\alpha}}$ dimensionless.
It is convenient to represent the sum of the contributions of all the species
of the fluid in the form
\be
\ve \equiv \ve^{\vphu}_{\mbsu{m}}(A) = \frac{3\lambda_{\mbsu{g}}}{\kappa_{\mbsu{g}}}\,
\mu(A) A^{-3},
\label{epsadep}
\ee
where
\be
\mu(A) = \sum_{\alpha\,\in\,[0,1]} \mu_{\dbss{\alpha}} A^{-\alpha}.
\label{def:mua}
\ee

For the dual matter governed by the metric
\be
\gtld_{\dbss{\mu\nu}} = \mbox{diag}\,\{-A^{-2},-A^{-2},-A^{-2},\,B^{-2}\}\,,
\label{def:tgmncl}
\ee
the energy-momentum tensor has the form analogous to (\ref{def:tpf}), (\ref{ucomfr}):
\be
\tilde{T}^{\mu}_{\nu} = (\tilde{\ve} + \tilde{p})\,
\tilde{u}^{\mu} \tilde{u}_{\dbss{\nu}} - \tilde{p}\,\delta^{\mu}_{\nu}\,,
\label{def:ttpf}
\ee
where
\be
\tilde{u}^k = 0, \quad k=1,2,3\,;\qquad  \tilde{u}^{4} \tilde{u}_{\dbss{4}} = 1\,.
\label{tucomfr}
\ee
The equation of state for the species also has the same form as (\ref{peeom}):
\be
\tilde{p} = \frac{\tilde{\alpha}}{3}\,\tilde{\ve}\,,\qquad \tilde{\alpha}\,\in\,[0,1]\,.
\label{tpeeom}
\ee
But the conservation law (\ref{ttmncons}) gives the opposite sign in the second term
of the right-hand side (r.h.s.) of the second relation between
$\tilde{p}$ and $\tilde{\ve}$ as compared to Eq.~(\ref{perelat})
\be
\tilde{p}(A) = -\tilde{\ve}(A) + \frac{A}{3}\,\frac{d\tilde{\ve}(A)}{dA}\,.
\label{tperelat}
\ee
This leads to the following result for
the sum of the contributions of all the species
\be
\tilde{\ve} \equiv \tilde{\ve}^{\vphu}_{\mbsu{m}}(A)
= \frac{3\lambda_{\mbsu{g}}}{\kappa_{\mbsu{g}}}\,\tilde{\mu}(A) A^{3},
\label{tepsadep}
\ee
where
\be
\tilde{\mu}(A) = \sum_{\tilde{\alpha}\,\in\,[0,1]}
\tilde{\mu}_{\dbss{\tilde{\alpha}}} A^{\tilde{\alpha}}.
\label{def:tmua}
\ee

\subsection{TDR equations: general case
\label{ssec:cl2}}

In this section the TDR equations are taken in the asymmetric form (\ref{eomas}).
Only time-like components of these equations
will be considered, because their space-like part can be obtained
in the given case by simple differentiation of the time-like components with respect to $x^4$.
Notice that in the model under consideration, the energy-momentum tensor density
$\mft^{\mu\nu}$ does not depend on the space-like coordinates, while its component $\mft^{44}$,
as follows from Eq. (\ref{taucons}), satisfies the equation $\partial_{\dbss{4}} \mft^{44} = 0$.
However, the sign of $\mft^{44}$ is changed under a time inversion
[see the remark after Eq. (\ref{actdc})], so it is not a genuine constant.
Such a constant can be constructed by multiplying $\mft^{44}$
by $\mbox{sgn}(B)$ where $B$ is the lapse function which is also the time-odd quantity.
Thus, the following equalities are fulfilled
\be
\ve_{\mbss{tot}} = \mbox{sgn}(B)\,\mft^{44}
= \frac{6\lambda_{\mbsu{g}}}{\kappa_{\mbsu{g}}}\,\tau_{\mbss{c}} = \mbox{constant},
\label{def:etot}
\ee
where $\ve_{\mbss{tot}}$ is the total energy density of the system ``matter plus gravitation''
in the cosmological limit, and the dimensionless constant $\tau_{\mbss{c}}$ is the value
of $\ve_{\mbss{tot}}$ expressed in units of
$6\lambda_{\mbsu{g}}/\kappa_{\mbsu{g}} = 3 \hbar c/(32\pi\,l^{(0)2}_{\mbsu{g}}\,l^2_{\mbsu{P}})$.

With the above definitions and remarks one has
\be
\mfe_{\zeta}\,G^{(\zeta)4}_{\,4} =
\frac{3 A^{\vphu}_{\zeta}\bigl(\partial_{\dbss{4}}A^{\vphu}_{\zeta}\bigr)^2}
{B^{\vphu}_{\zeta}},
\label{def:ezgzt}
\ee
where
\be
A^{\vphu}_{\zeta} = A + \zeta\,A^{-1}, \qquad
B^{\vphu}_{\zeta} = B + \zeta\,B^{-1},
\label{def:azbz}
\ee
\be
\partial_{\dbss{4}}A^{\vphu}_{\zeta}
= \bigl(1 - \zeta\,A^{-2}\bigr)\,\partial_{\dbss{4}}A\,,
\label{def:d4az}
\ee
and
\begin{subequations}
\bea
\mfe_{\zeta}\,Y^4_4 &=& -3\lambda_{\mbsu{g}}\,\bigl[ B\,w + \mbox{sgn}(B)\,\tau_{\mbss{c}}\bigr]\,,
\label{ezy44}\\
\mfe_{\zeta}\,Z^4_4 &=& -3\lambda_{\mbsu{g}}\,\bigl[ B^{-1}\tilde{w}
+ \mbox{sgn}(B)\,\tau_{\mbss{c}}\bigr]\,,
\label{ezz44}
\eea
\label{ezyz44}
\end{subequations}
where
\begin{subequations}
\bea
w &=& A^{-3}(A^2 - 1)(3 A^2 - 1) - \mu(A)\,,
\label{def:wa}\\
\tilde{w} &=& A^{-1}(A^2 - 1)(A^2 - 3) - \tilde{\mu}(A)\,.
\label{def:twa}
\eea
\label{def:watwa}
\end{subequations}

Substitution of Eqs. (\ref{ezyz44}) into Eq. (\ref{eomas2}) leads to the equation
\be
\tau_{\mbss{c}} B^2 + \tilde{w}^{\vphu}_{\zeta}|B| - \zeta \tau_{\mbss{c}} = 0\,,
\label{b2eq}
\ee
where
\be
\tilde{w}^{\vphu}_{\zeta} = \tilde{w} - \zeta w\,.
\label{def:twz}
\ee
Equation (\ref{b2eq}) has two solutions, but the condition of positivity
of $|B|$ at $\zeta > 0$ retains only one of them:
\be
|B| = |B(A)| = \frac{\mbox{sgn}(\tau_{\mbss{c}})\sqrt{\tilde{w}^{2}_{\zeta}
+ 4\zeta \tau_{\mbsu{c}}^2} - \tilde{w}^{\vphu}_{\zeta}}{2 \tau_{\mbss{c}}}\,.
\label{b2sol}
\ee

At this stage it makes sense
to perform the change of variable $x^4 \rightarrow t$ determined by the equation
\be
dt(x^4)/dx^4 = B(A(x^4))/c\,,
\label{tran:x4t}
\ee
where $c$ is the speed of light,
that brings the flat FRW metric (\ref{def:gmncl})
to the form mostly used in applications
\be
g^{\mbsu{FRW}}_{\mu\nu} = \mbox{diag}\,\{-A^2,-A^2,-A^2,\,c^2\}\,.
\label{def:gFRW}
\ee
This transformation also gives
\be
\partial_{\dbss{4}}A = c^{-1}B\dot{A}\,,\qquad \dot{A} = dA/dt\,.
\label{tran:dA}
\ee
Now, substituting the formulas (\ref{def:ezgzt})--(\ref{ezyz44})
into Eq.~(\ref{eomas1}) one obtains with account of (\ref{tran:dA})
the equation
\be
\dot{A}^2 + {\mathcal U}_{\mbss{c}}(A) = 0\,,
\label{baseqcl}
\ee
where the function ${\mathcal U}_{\mbss{c}}(A)$ can be called
a cosmological quasipotential which is defined as
\be
{\mathcal U}_{\mbss{c}}(A) = \frac{c^2\lambda_{\mbsu{g}} (1+\zeta B^{-2})^2}
{(\lindc{1}+\zeta A^{-2})(\lindc{1}-\zeta A^{-2})^2}
\left(\,\frac{w}{A} + \frac{\tau_{\mbss{c}}}{A|B|}\,\right)
\label{def:vcosm}
\ee
with $|B|$ and $w$ determined by Eqs. (\ref{def:watwa}), (\ref{def:twz}),
and (\ref{b2sol}).
%

As can be seen from Eqs. (\ref{b2sol}) and (\ref{def:vcosm}), the non-zero
and positive value of the parameter $\zeta$ enables one to remove singularities
of the function ${\mathcal U}_{\mbss{c}}(A)$ which arise in the case $\zeta = 0$
in the points where $\tilde{w}(A)=0$.
If one neglects the contribution of $\tilde{\mu}(A)$ into the r.h.s. of
Eq.~(\ref{def:twa}), one obtains two such points at $A>0:$ $A=1$ and $A=\sqrt{3}$.
The contribution of $\tilde{\mu}(A)$ shifts the first point to the left and
the second point to the right that leads to the broadening of the interval
between them.

\subsection{Positive energy density and stable Universe
\label{ssec:cl3}}

Consider the case of the positive total energy density $\tau_{\mbss{c}}$
in Eqs. (\ref{baseqcl}) and (\ref{def:vcosm}).
If $\tau_{\mbss{c}}>0$, the term $\tau_{\mbss{c}}/(A|B|)$
in Eq. (\ref{def:vcosm}) is everywhere positive, and one can try to find
the conditions under which the quasipotential ${\mathcal U}_{\mbss{c}}(A)$
is also everywhere positive except for the one point $A=A_{\mbsu{st}}$
where it is equal to zero. If such conditions are fulfilled,
one obtains the stable solution of Eq.~(\ref{baseqcl}):
$A=\mbox{constant}=A_{\mbsu{st}}$.

It is natural to demand that this stable solution
corresponds to the background metric $\gamma_{\dbss{\mu\nu}}$ that enters
the equilibrium conditions (\ref{eqmet}). This requirement is formulated
with the help of the following equations
\begin{subequations}
\bea
|B(1)| &=& 1\,,
\label{stcond1}\\
{\mathcal U}_{\mbss{c}}(1) &=& 0\,,
\label{stcond2}\\
{\mathcal U}^{\prime}_{\mbsu{c}}(1) &=& 0\,.
\label{stcond3}
\eea
\label{stcond}
\end{subequations}
The additional condition of stability
\be
{\mathcal U}^{\prime\prime}_{\mbsu{c}}(1) > 0
\label{stcond4}
\ee
should be also fulfilled but it should be a consequence
of Eqs.~(\ref{stcond}).

\begin{figure}[]
\begin{center}
\includegraphics*[trim=2cm 15cm 0cm 2cm,clip=true,scale=0.5,angle=0]{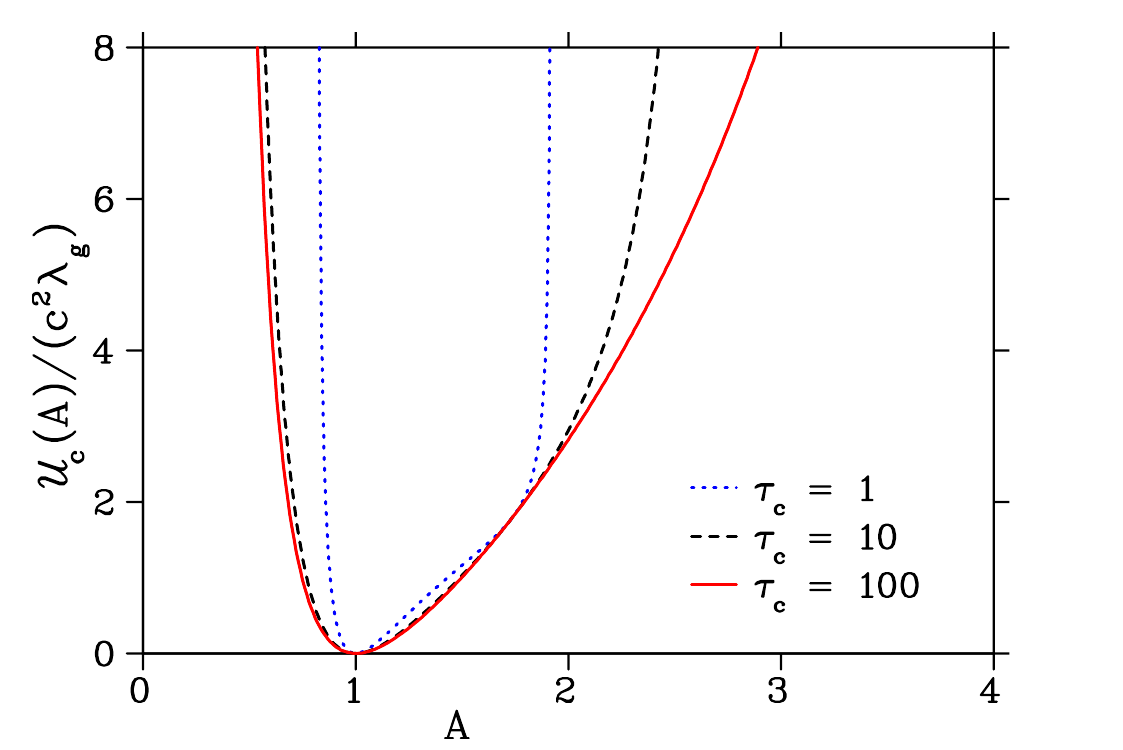}
\end{center}
\caption{\label{fig:vst}
Cosmological quasipotential ${\mathcal U}_{\mbss{c}}(A)$ in the vicinity
of the equilibrium point $A=1$ in the case of the fulfillment
of the conditions (\ref{emtmc}).
}
\end{figure}

It can be verified that the equations (\ref{stcond})
for the functions $|B(A)|$ and ${\mathcal U}_{\mbss{c}}(A)$
are fulfilled (at any $\zeta$) if the following conditions are satisfied
\begin{subequations}
\bea
&&\tau_{\mbss{c}} = \mu(1) = \tilde{\mu}(1)\,,
\label{emtm1}\\
&&\mu^{\prime}(1) = - \tilde{\mu}^{\prime}(1)\,.
\label{dmdtm1}
\eea
\label{emtmc}
\end{subequations}
In this case for the second derivative of ${\mathcal U}_{\mbss{c}}(A)$
one has
\bea
&&\hspace{-1em}
{\mathcal U}^{\prime\prime}_{\mbsu{c}}(1) =
\frac{t^{-2}_{\mbsu{g}}}{8}
\nonumber\\
&&\hspace{-1em}
\times\Bigl\{ 16 + 2\bigl[4 - \mu^{\prime}(1)\bigr]^2/\mu(1)
- \mu^{\prime\prime}(1) - \tilde{\mu}^{\prime\prime}(1) \Bigr\}\,,
\label{ddvc1}
\eea
where
\be
t^{-1}_{\mbsu{g}} = m^{(\zeta)}_{\mbsu{g}}c^2/\hbar
\label{def:tg}
\ee
is the inverse Compton time of the graviton, see Eq.~(\ref{gravmass}).
From this formula it is seen that the condition (\ref{stcond4})
is also fulfilled at the positive $\lambda_{\mbsu{g}}$ if the values of
the second derivatives of the functions $\mu(A)$ and $\tilde{\mu}(A)$
in the point $A=1$ are sufficiently small (at least, it is always fulfilled
in the case of the dust-like matter for which $\mu(A)$ and $\tilde{\mu}(A)$
are the constants).
From Eqs. (\ref{emtmc}) and (\ref{ddvc1}) it follows also that
the existence of both ordinary and dual matter is necessary for the fulfillment
of the stability conditions (\ref{stcond}) and (\ref{stcond4}).
The shape of the cosmological quasipotential in the vicinity of the equilibrium
point $A=1$ in the case of the fulfillment of the conditions (\ref{emtmc})
is shown in Fig.~\ref{fig:vst}. The approximation of the dust-like matter
and the limit $\zeta \rightarrow +0$ are assumed.

Thus, the TDR enables one to formulate the model of the stable Universe
which is described by the solution of Eq.~(\ref{baseqcl}) under
the conditions (\ref{emtmc}).
The averaged physical metrics of such a Universe,
$g_{\dbss{\mu\nu}}$ and $\gtld_{\dbss{\mu\nu}}$, coincide with the background
metric $\gamma_{\dbss{\mu\nu}}$. It means that the averaged gravitational field
of the stable Universe vanishes, that formally is expressed by
Eqs. (\ref{eqmet}), (\ref{eqsol1}), and (\ref{eqcond}).

\subsection{Negative energy density\\
and oscillating cosmological scale factor
\label{ssec:cl4}}

\subsubsection{Hubble parameter and the description of data
\label{sssec:hp}}

Suppose that in the stable Universe with positive total energy density $\tau_{\mbss{c}}$,
there exist the domains where $\tau_{\mbss{c}}<0$ and that the sizes of these domains
can be very large (comparable to or greater than the size of visible Universe).
As will be shown below, the present-day state of our part of the Universe
and the available data of measurements of the Hubble parameter
can be described by the solution of Eq.~(\ref{baseqcl}) in such a domain at the values
of the scale factor $A$ in the range from 0 to 1.
Let $t^{\vphu}_0$ be the current value of the time variable (the present time)
and $A^{\vphu}_0 = A(t^{\vphu}_0)$. It is convenient to introduce the
reduced scale factor $a(t)$ as
\be
A(t) = A^{\vphu}_0\,a(t)\,.
\label{def:redsf}
\ee
Then, according to Eq.~(\ref{baseqcl}), the square of the Hubble parameter $H(a(t))$
is defined by the equation
\be
H^2(a) = \dot{a}^2/a^2 = - {\mathcal U}_{\mbss{c}}(A^{\vpsq}_0 a)/(A^{\vpsq}_0 a)^2.
\label{def:Hpar}
\ee

From the standpoint of analysis of the data,
the realistic situation is the case when one can set the mixing parameter
$\zeta$ equal to zero (more general cases will be considered in the
subsection \ref{sssec:cqp}). Then, assuming $\zeta = 0$, one can represent
the function $H^2(a)$ defined by Eqs. (\ref{def:vcosm}) and (\ref{def:Hpar})
in the form
\be
H^2(a) = H^2_0\left(\Omega_{\mu}a^{-3}
+ \Omega_{\lambda}\frac{F_{\lambda}(a)}{F_{\lambda}(1)}
+ \Omega^{\vphu}_{\tau}\frac{F_{\tau}(1)}{F_{\tau}(a)}\right),
\label{def:H2a}
\ee
where
\begin{subequations}
\bea
H^2_0 &=& H^2(1)\,,
\label{def:H20}\\
H^2_0\,\Omega_{\mu} &=& c^2\lambda_{\mbsu{g}}\,A^{-3}_0\mu^{\vphu}_0\,,
\label{def:omgm}\\
H^2_0\,\Omega_{\lambda} &=& c^2\lambda_{\mbsu{g}}\,F_{\lambda}(1)\,,
\label{def:omgl}\\
H^2_0\,\Omega^{\vphu}_{\tau} &=& c^2\lambda_{\mbss{g}}\,\tau_{\mbsu{c}}^2/F_{\tau}(1)\,,
\label{def:omge}
\eea
\label{def:omg}
\end{subequations}
\begin{subequations}
\bea
F_{\lambda}(a) &=& (A^{\vphu}_0 a)^{-6}\,\bigl[1 - (A^{\vphu}_0 a)^2\bigr]\,
\bigl[3(A^{\vphu}_0 a)^2 - 1\bigr]\,,\hspace{3em}
\label{def:Flam}\\
F_{\tau}(a) &=& (A^{\vphu}_0 a)^2\,\bigl[1 - (A^{\vphu}_0 a)^2\bigr]\,
\bigl[3 - (A^{\vphu}_0 a)^2\bigr]
\nonumber\\
&-& (A^{\vphu}_0 a)^3 \tilde{\mu}^{\vphu}_0.
\label{def:Feps}
\eea
\label{def:Fle}
\end{subequations}
Here and in what follows the model of the dust-like matter is used
in which $\mu(A) = \mu^{\vphu}_0$ and $\tilde{\mu}(A) = \tilde{\mu}^{\vphu}_0$,
where $\mu^{\vphu}_0$ and $\tilde{\mu}^{\vphu}_0$ are the constants.
The parameters $\Omega_{\mu}$, $\Omega_{\lambda}$, and $\Omega^{\vphu}_{\tau}$ are
related by the equality
\be
\Omega_{\mu} + \Omega_{\lambda} + \Omega^{\vphu}_{\tau} = 1\,,
\label{omgrel}
\ee
that obviously follows from the definitions (\ref{def:H2a}) and (\ref{def:H20}).

Formula (\ref{def:H2a}) contains 5 independent model parameters: $A^{\vphu}_0$,
$H^2_0\,\Omega_{\mu}$, $H^2_0\,\Omega_{\lambda}$, $H^2_0\,\Omega^{\vphu}_{\tau}$,
and $\tilde{\mu}^{\vphu}_0$, which in principle can be
determined by a least-squares fit to the available $H(z)$ data,
where $z = a^{-1} - 1$ is the redshift.
To this end, the following function of the model parameters was minimized
\be
\chi^2 = \sum^{N_{\mbtu{D}}}_{i=1} \chi^2_i\,,\qquad
\chi^{\vpb}_i = \frac{H^2_{i\,(\mbsu{th})} - H^2_{i\,(\mbsu{obs})}}
{2\,H_{i\,(\mbsu{obs})} \Delta H_{i\,(\mbsu{obs})}}\,,
\label{def:chi2}
\ee
where $H_{i\,(\mbsu{obs})} = H(z^{\vpb}_i)$ is the observed value
of $H(z)$ at the redshift $z = z^{\vpb}_i$, $\;\Delta H_{i\,(\mbsu{obs})}$ is
its uncertainty, and $N^{\vphu}_{\mbsu{D}}$ is the number of data points used in the fit.
$H^2_{i\,(\mbsu{th})} = H^2(1/(1+z^{\vpb}_i))$ is the square of the Hubble parameter
computed within a certain theoretical model
[determined by Eqs. (\ref{def:H2a})--(\ref{def:Fle}) in the case of the TDR].
For the sake of simplicity, in the present work the parameter $\tilde{\mu}^{\vphu}_0$
was set equal to zero. The remaining 4 parameters of Eqs. (\ref{def:H2a})--(\ref{def:Fle})
have been found using the data from Refs. \cite{Farooq17,Yu18,Cao23}.

\begin{table*}[]
\caption{\label{tab:1}
Parameters of the function $H^2(a)$ in the TDR, Eqs. (\ref{def:H2a})--(\ref{def:Fle}),
and in the $\Lambda$CDM model, Eq. (\ref{def:H2LCDM}), found from the fit
to the $H(z)$ data from Refs. \cite{Farooq17,Yu18,Cao23}.
The other quantities are defined in the text.
}
\begin{ruledtabular}
\begin{tabular}{lddddddd}
$H(z)$ data reference & \multicolumn{5}{c}{$\hspace{1.5em}$ \cite{Farooq17} }
& \multicolumn{1}{c}{$\hspace{1em}$ \cite{Yu18}} & \multicolumn{1}{c}{$\hspace{1em}$ \cite{Cao23}} \\
\cline{2-6}
& \multicolumn{1}{r}{WM34} & \multicolumn{1}{r}{WM45} & \multicolumn{1}{c}{$\hspace{1em}\lindc$ WM57}
& \multicolumn{1}{r}{WM456} & \multicolumn{1}{r}{Full set} && \\
\hline
TDR: $\lindc$ &&&&&&& \\
$A^{\vphu}_0$
& 0.8766 & 0.9098 & 0.9938 & 0.9564 & 0.8404 & 0.8345 & 0.8740 \\
$\Omega_{\mu}$
& 0.2102 & 0.2074 & 0.0290 & 0.1308 & 0.2072 & 0.2295 & 0.2056 \\
$\Omega_{\lambda} \times 10^{4}$
& 5.039 & 5.119 & 0.08339 & 1.9847 & 4.356 & 5.830 & 8.726 \\
$\Omega^{\vphu}_{\tau}$
& 0.7893 & 0.7921 & 0.9710 & 0.8690 & 0.7923 & 0.7699 & 0.7935 \\
$H^{\vphu}_0$ (km s$^{-1}$Mpc$^{-1}$)
& 80.08 & 84.82 & 240.9 & 108.6 & 76.52 & 75.17 & 80.07 \\
$q^{\vphu}_0$
& -2.784 & -4.004 & -77.87 & -9.631 & -2.045 & -1.882 & -2.733 \\
$j^{\vphu}_0 \times 10^{-2}$
& 0.4185 & 0.8380 & 249.1 & 4.264 & 0.2334 & 0.2084 & 0.4017 \\
$\mu^{\vphu}_0$
& 187.2 & 137.4 & 86.02 & 112.1 & 263.4 & 224.0 & 107.6 \\
$\tau_{\mbss{c}}$
& -20.35 & -14.69 & -8.508 & -11.77 & -28.41 & -25.10 & -15.84 \\
$\ve^{\vphu}_{\mbsu{m}}(A^{\vphu}_0)$ (GeV m$^{-3}$)
& 1.421 & 1.572 & 1.770 & 1.624 & 1.278 & 1.367 & 1.389 \\
$\ve_{\mbss{tot}}\vphantom{\sum_{S_Q}}$ (MeV m$^{-3}$)
& -208.1 & -253.2 & -343.7 & -298.4 & -163.7 & -177.9 & -273.0 \\
$(m^{\vphu}_{\mbsu{g}}/m^{\vphu}_{\mbsu{H}}) \times 10^{2}$
& 7.779 & 9.536 & 5.141 & 9.037 & 6.113 & 6.902 & 10.10 \\
$(m^{\vphu}_{\mbsu{g}}/m^{\vphu}_{\mbsu{P}}) \times 10^{62}$
& 1.088 & 1.413 & 2.164 & 1.714 & 0.8172 & 0.9064 & 1.413 \\
$l^{\vphu}_{\mbsu{g}}$ (Gpc)
& 48.12 & 37.06 & 24.21 & 30.56 & 64.09 & 57.79 & 37.06 \\
$(A^2_0/B^2_0) \times 10^{-3}$
& 0.9162 & 1.057 & 113.6 & 3.662 & 0.8874 & 0.6244 & 0.5254 \\
\hline
$\Lambda$CDM: $\lindc$ &&&&&&& \\
$\Omega_{\mbss{m}}$
& 0.2625 & 0.2644 & 0.2641 & 0.2613 & 0.2610 & 0.2609 & 0.2938 \\
$\Omega_{\Lambda}$
& 0.7375 & 0.7356 & 0.7359 & 0.7387 & 0.7390 & 0.7391 & 0.7062 \\
$H^{\vphu}_0$ (km s$^{-1}$Mpc$^{-1}$)
& 69.78 & 69.70 & 69.76 & 69.91 & 69.78 & 70.27 & 68.48 \\
$q^{\vphu}_0$
& -0.6063 & -0.6034 & -0.6039 & -0.6080 & -0.6085 & -0.6087 & -0.5593 \\
$\ve^{\vphu}_{\mbsu{m}}$ (GeV m$^{-3}$)
& 1.347 & 1.354 & 1.354 & 1.346 & 1.339 & 1.357 & 1.451 \\
$\lindc{\chi^2_{\mbsu{$\Lambda$CDM}}}/\chi^2_{\mbsu{TDR}}$
& 1.64 & 3.69 & 51.5 & 9.43 & 1.12 & 1.21 & 1.02 \\
$N^{\vphu}_{\mbsu{D}}$
& 12 & 10 & 8 & 8 & 38 & 36 & 32 \\
\end{tabular}
\end{ruledtabular}
\end{table*}

The results are listed in Table~\ref{tab:1}.
The columns entitled ``WM34'', ``WM45'', ``WM57'', and ``WM456'' refer to the binned
$H(z)$ data from Ref.~\cite{Farooq17}.
In this table,
the quantities $\mu^{\vphu}_0$, $\tau_{\mbss{c}}$, $l^{\vphu}_{\mbsu{g}}$,
$\ve^{\vphu}_{\mbsu{m}}(A^{\vphu}_0)$, and $\ve_{\mbss{tot}}$
are determined with the use of the computed parameters of the function (\ref{def:H2a})
by the equations (\ref{gravcwl}), (\ref{epsadep}), (\ref{def:mua}), (\ref{def:etot}),
and (\ref{def:omg}).
The ratios of the graviton mass $m^{\vphu}_{\mbsu{g}}$ to the Hubble mass $m^{\vphu}_{\mbsu{H}}$
(with $m^{\vphu}_{\mbsu{H}}c^2 = \hbar H^{\vphu}_0$) and to the Planck mass $m^{\vphu}_{\mbsu{P}}$
are determined with the use of Eqs.~(\ref{gravmass}) [here the superindex ``(0)''
in the notations $l^{(0)}_{\mbsu{g}}$ and $m^{(0)}_{\mbsu{g}}$ is dropped, supposing
the smallness of the parameter $\zeta$].
The ratios of the squares of the present-day values of the scale factor and of the lapse function
$A^2_0/B^2_0$ listed in the table will be discussed in Sec.~\ref{sssec:pdm}.
The present-day values of the deceleration parameter $q^{\vphu}_0 = q(1)$ and
jerk parameter $j^{\vphu}_0 = j(1)$ were defined by the formulas
\be
q(a) = - \frac{\bigl(a^2 H^2(a)\bigr)^{\prime}}{2aH^2(a)},\qquad
j(a) = \frac{\bigl(a^2 H^2(a)\bigr)^{\prime\prime}}{2H^2(a)}
\label{def:qaja}
\ee
with the function $H^2(a)$ from Eq.~(\ref{def:H2a}).

The lower part of Table~\ref{tab:1} contains the results of the fit
for the flat $\Lambda$CDM model \cite{Peebles84}.
In this model the function $H^2(a)$ has the form
\be
H^2(a) = H^2_0\left(\,\Omega_{\mbss{m}} a^{-3} + \Omega_{\Lambda}\,\right)
\label{def:H2LCDM}
\ee
with $\Omega_{\mbss{m}} + \Omega_{\Lambda} = 1$ and $j^{\vphu}_0 = 1$.
In the case of the $\Lambda$CDM, the energy density of the matter
$\ve^{\vphu}_{\mbsu{m}}$ was defined by the formula
$\ve^{\vphu}_{\mbsu{m}} = \Omega_{\mbss{m}} \rho^{\vphu}_{\mbsu{crit}}c^2$
where $\rho^{\vphu}_{\mbsu{crit}} = 3H^2_0/(c^4 \kappa_{\mbsu{g}})$
is the conventional critical density \cite{Misner73,Landau71}.
Thus determined quantity $\ve^{\vphu}_{\mbsu{m}}$ should be compared with
the analogous energy density $\ve^{\vphu}_{\mbsu{m}}(A^{\vphu}_0)$ in the TDR
[however, $\ve^{\vphu}_{\mbsu{m}}(A^{\vphu}_0)$ cannot be considered
as a certain contribution of the ordinary matter into the total energy density
$\ve_{\mbss{tot}}$, because this contribution has no an absolute meaning,
see Sec.~\ref{ssec:cl}].
As can be seen, both quantities, $\ve^{\vphu}_{\mbsu{m}}$ and
$\ve^{\vphu}_{\mbsu{m}}(A^{\vphu}_0)$,
are close in magnitude in spite of the rather large
spread in the values of other parameters.

\begin{figure}[]
\begin{center}
\includegraphics*[trim=2cm 14.6cm 0cm 2cm,clip=true,scale=0.46,angle=0]{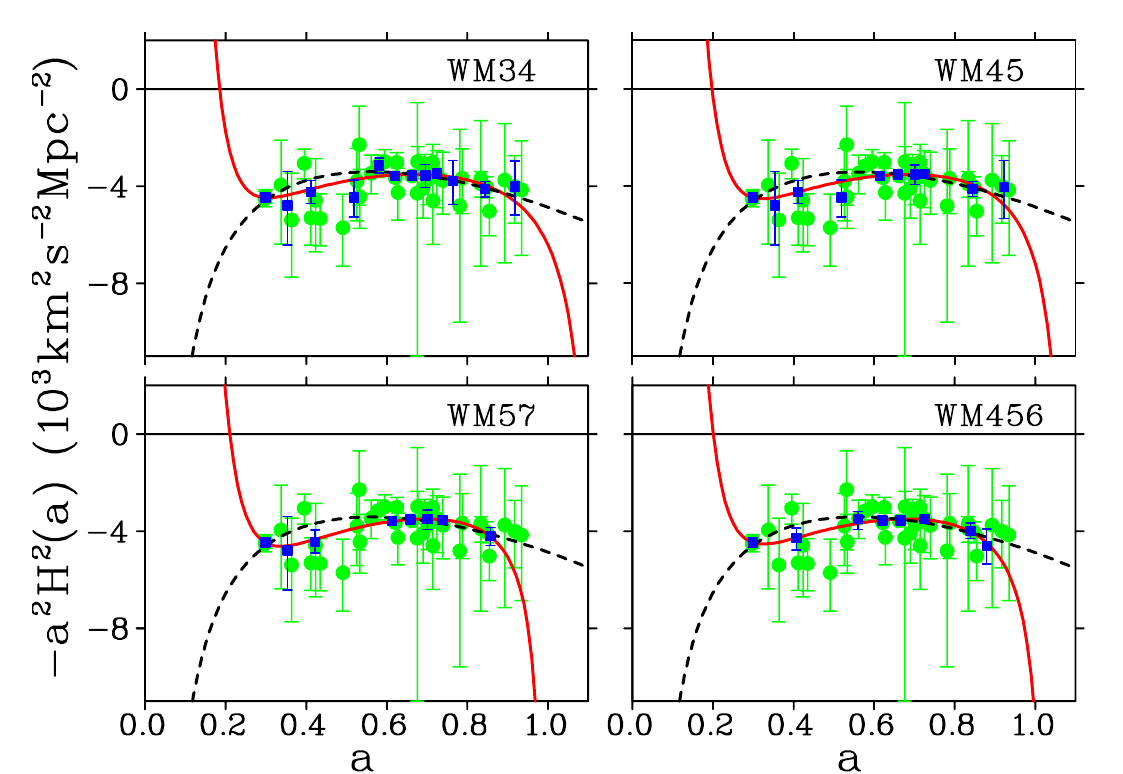}
\end{center}
\caption{\label{fig:fit}
The function of the Hubble parameter $-a^2 H^2(a)$ in comparison to the
data from Ref.~\cite{Farooq17}. The binned data sets WM34, WM45, WM57, and WM456
are shown by the blue squares. The green circles represent the full set of data.
The red solid and black dashed lines correspond to the functions calculated within
the TDR and $\Lambda$CDM, respectively.
}
\end{figure}

The last but one row of Table~\ref{tab:1} contains the ratios of the
$\chi^2$-functions (\ref{def:chi2}) calculated with the use of
the $\Lambda$CDM model and of the TDR.
The TDR provides a better description of the data in all the cases,
but especially it refers to the binned data. In particular,
$\chi^2_{\mbsu{$\Lambda$CDM}}/\chi^2_{\mbsu{TDR}} = 51.5$ and 9.4
in the cases of the WM57 and WM456 sets, respectively.

The obtained results are illustrated in Fig.~\ref{fig:fit} where the $a$-dependence
of the function $-a^2 H^2(a)$ is shown for the TDR and $\Lambda$CDM
in comparison to the respectively transformed $H(z)$ data from Ref.~\cite{Farooq17}.
As follows from Eq.~(\ref{def:Hpar}), this function is the scaled
cosmological quasipotential which will be discussed in more detail in the next subsection.
The quality of description of the full set of data from Ref.~\cite{Farooq17}
and of the data from \cite{Yu18,Cao23}
is approximately the same in the TDR and $\Lambda$CDM,
mainly because of the large data dispersion.
However, this is not the case for the binned data for which the TDR results
look much better in agreement with the values of $\chi^2$ ratios from Table~\ref{tab:1}.
In particular, visually the TDR provides nearly perfect coincidence with data
for the WM57 and WM456 sets.
It is also important to note that the binned data show a tendency to the upward bend
of the function $-a^2 H^2(a)$ at $a \lesssim 0.3$ which is most definitely manifested
in the case of the WM57 set.
This bend is described in the TDR but is absent in the $\Lambda$CDM model.

\subsubsection{Cosmological quasipotential and its properties
\label{sssec:cqp}}

In the majority of the cosmological models based on the GR,
in particular in the $\Lambda$CDM,
the behavior of the quasipotential near the limiting values of the scale factor
is singular.
In these models, the quasipotential determined as the function
$-a^2 H^2(a)$ goes to $-\infty$ both at $a \rightarrow +0$ and at $a \rightarrow +\infty$.
In the case of the flat $\Lambda$CDM model, it is readily seen from Eq.~(\ref{def:H2LCDM}).
Thus, the ``motion'' of the scale factor is here infinite.
The zero-point singularity of such time evolution is usually interpreted
in terms of the Big Bang model \cite{Overduin04,Ryden17}.

In the TDR, the cosmological quasipotential ${\mathcal U}_{\mbss{c}}(A)$
is defined by Eqs. (\ref{def:watwa}), (\ref{def:twz}), (\ref{b2sol}),
and (\ref{def:vcosm}). In the case of $\tau_{\mbss{c}}<0$,
its behavior at $A \rightarrow +0$ and $A \rightarrow +\infty$ is determined
in the limit $\zeta \rightarrow +0$ by the function $w(A)/A$.
As follows from Eqs. (\ref{def:mua}) and (\ref{def:wa}),
$w(A)A^3 \rightarrow 1$ at $A \rightarrow +0$ and $w(A)/A \rightarrow 3$
at $A \rightarrow +\infty$.
So, the function $w(A)/A$ and consequently the function
${\mathcal U}_{\mbss{c}}(A)$ are positive in both cases.
It means that the time evolution of the scale factor $A$ in the TDR
has a character of oscillations within the finite limits excluding the zero point
that is incompatible with the standard scenario of the Big-Bang cosmology.
The opposite behavior of the quasipotentials in the TDR and $\Lambda$CDM near $a=0$
is illustrated in Fig.~\ref{fig:fit}.

However, in the TDR, the function ${\mathcal U}_{\mbss{c}}(A)$
can have the quasisingular character in the another range of values of $A$.
Consider the limit $\zeta \rightarrow +0$ and $\tilde{\mu} = 0$.
In this case (and at $\tau_{\mbss{c}}<0$),
according to Eqs. (\ref{def:twa}), (\ref{def:twz}), and (\ref{b2sol}),
$|B| = -\tilde{w}/\tau_{\mbss{c}}$ at $0< A < 1$ and
$A > \sqrt{3}$, where $\tilde{w} > 0$, and $|B| \rightarrow +0$
at $1 < A < \sqrt{3}$, where $\tilde{w} < 0$.
In the latter case, ${\mathcal U}_{\mbss{c}}(A) \rightarrow -\infty$
that leads to the sharp (instantaneous at $\zeta \rightarrow +0$)
jump of the scale factor $A$ from 1 to $\sqrt{3}$ in the course of its increase
or to the drop from $\sqrt{3}$ to 1 in the course of its decrease.
As was mentioned in the end of subsection~\ref{ssec:cl2},
the contribution of the energy density of the dual matter $\tilde{\mu}(A)$
leads to the broadening of the interval of the values of $A$,
in which $\tilde{w}(A)$ is negative, but it does not change the general picture.

\begin{figure}[]
\begin{center}
\includegraphics*[trim=2cm 16cm 0cm 2cm,clip=true,scale=0.5,angle=0]{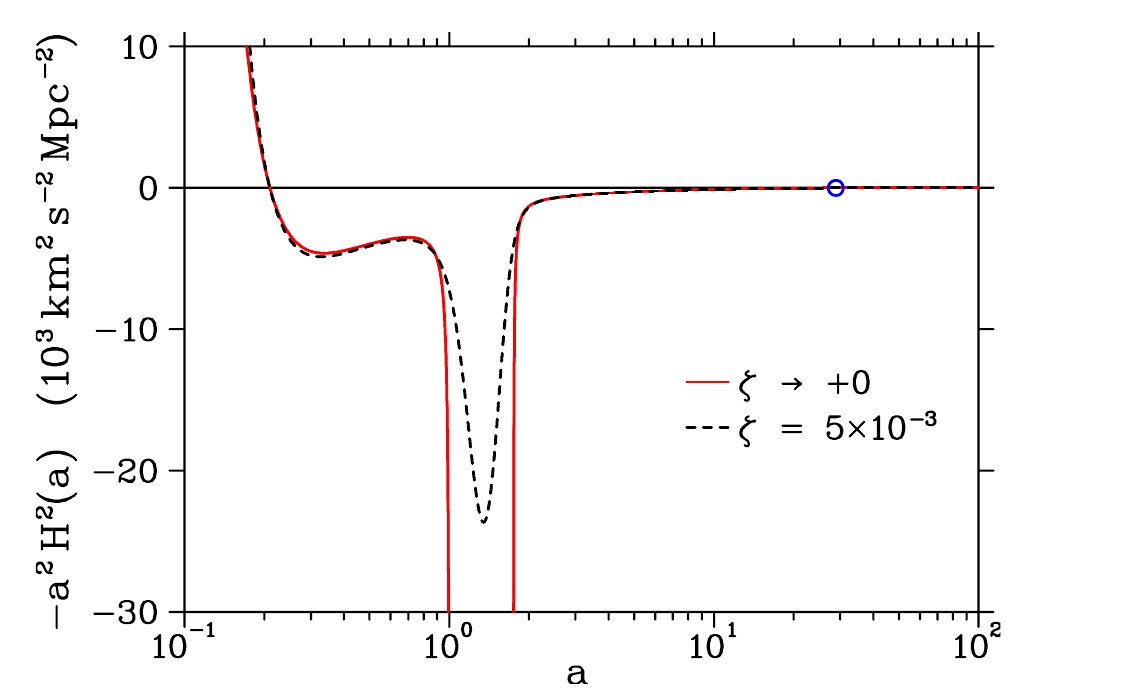}
\end{center}
\caption{\label{fig:wm57}
Scaled cosmological quasipotential with parameters fitted to the binned data set WM57
from Ref.~\cite{Farooq17}.
}
\end{figure}

Thus, in the sufficiently general case of interest, the cosmological quasipotential
${\mathcal U}_{\mbss{c}}(A)$ in the TDR
at the negative total energy density and the small value
of the mixing parameter $\zeta$ has four characteristic points:
two turning points (left point $A_{\mbsu{t.L}}$ and right point $A_{\mbsu{t.R}}$)
in which ${\mathcal U}_{\mbss{c}}(A)=0$ and
two critical points (left point $A_{\mbsu{c.L}}$ and right point $A_{\mbsu{c.R}}$)
in which the sign of the function $\tilde{w}(A)$ defined in Eq.~(\ref{def:twa}) is changed.

This form of the cosmological quasipotential is demonstrated in Fig.~\ref{fig:wm57} where
it is represented by the function $-a^2 H^2(a)$ for $H^2(a)$
with parameters fitted to the binned data set WM57 from Ref.~\cite{Farooq17}.
In the case of $\zeta \rightarrow +0$ (red solid line) the quasipotential goes
to $-\infty$ in the region between the critical points\\
$a_{\mbsu{c.L}} = A_{\mbsu{c.L}}/A^{\vphu}_0 = 1/A^{\vphu}_0 \approx 1.01$
and\\
$a_{\mbsu{c.R}} = A_{\mbsu{c.R}}/A^{\vphu}_0 = \sqrt{3}/A^{\vphu}_0
\approx 1.74$.\\
Behavior of the quasipotential in this region
at the finite (but small) values of the mixing parameter $\zeta$ is shown
by the black dashed line obtained in the calculation with $\zeta = 5\times 10^{-3}$.
In both cases, one has oscillations of the scale factor $a$
in the interval limited by the turning points\\
$a_{\mbsu{t.L}} = A_{\mbsu{t.L}}/A^{\vphu}_0 \approx 0.21$
and\\
$a_{\mbsu{t.R}} = A_{\mbsu{t.R}}/A^{\vphu}_0 \approx 28.90$\\
(the latter is indicated by the blue open circle in Fig.~\ref{fig:wm57}).

\begin{table*}[]
\caption{\label{tab:2}
The values of the turning points $a_{\mbsu{t.L}}$ and $a_{\mbsu{t.R}}$,
critical points $a_{\mbsu{c.L}}$ and $a_{\mbsu{c.R}}$,
and the time intervals of the oscillations of the scale factor
for the cosmological quasipotential with parameters given in Table~\ref{tab:1}.
See text for more details.
}
\begin{ruledtabular}
\begin{tabular}{lddddddd}
$H(z)$ data reference & \multicolumn{5}{c}{$\hspace{1.5em}$ \cite{Farooq17} }
& \multicolumn{1}{c}{$\hspace{1em}$ \cite{Yu18}} & \multicolumn{1}{c}{$\hspace{1em}$ \cite{Cao23}} \\
\cline{2-6}
& \multicolumn{1}{r}{WM34} & \multicolumn{1}{r}{WM45} & \multicolumn{1}{c}{$\hspace{1em}\lindc$ WM57}
& \multicolumn{1}{r}{WM456} & \multicolumn{1}{r}{Full set} && \\
\hline
$a_{\mbsu{t.L}}$
& 0.1850 & 0.1974 & 0.2101 & 0.2006 & 0.1721 & 0.1826 & 0.2179 \\
$a_{\mbsu{t.R}}$
& 71.21 & 50.36 & 28.90 & 39.10 & 104.5 & 89.50 & 41.09 \\
$a_{\mbsu{c.L}}$
& 1.141 & 1.099 & 1.006 & 1.046 & 1.190 & 1.198 & 1.144 \\
$a_{\mbsu{c.R}}$
& 1.976 & 1.904 & 1.743 & 1.811 & 2.061 & 2.076 & 1.982 \\
$T_{\mbsu{L}}$ (Gyr)
& 13.09 & 12.85 & 12.29 & 12.65 & 13.32 & 13.12 & 12.93 \\
$T_{\mbsu{C}}$ (Myr)
& 1193 & 793.1 & 17.01 & 280.5 & 1684 & 1801 & 1221 \\
$T^{\vphu}_{\zeta}\,\zeta^{-3/4}$ ($10^{18}$ s)
& 24.4 & 16.0 & 7.93 & 11.8 & 38.4 & 32.5 & 16.6 \\
$T^{\,(\mbsu{tent})}_{\zeta}$ ($10^{-28}$ s)
& 8.22 & 6.54 & 4.48 & 5.58 & 10.4 & 9.56 & 6.79 \\
$T_{\mbsu{R}}$ (Tyr)
& 25.05 & 14.14 & 5.763 & 9.500 & 46.99 & 36.02 & 11.05 \\
$T_{\mbsu{s.c.}}$ (Gyr)
& 27.37 & 26.50 & 24.59 & 25.57 & 28.32 & 28.05 & 27.09 \\
$T_{\mbsu{U}}$ (Gyr) [$\Lambda$CDM]
& 14.02 & 14.01 & 14.00 & 14.01 & 14.04 & 13.94 & 13.85 \\
\end{tabular}
\end{ruledtabular}
\end{table*}

\subsubsection{Time characteristics of the oscillations
\label{sssec:tco}}

One of the characteristics of the time evolution of the scale factor
is the time during which its value is changed in the given interval $(a_1,a_2)$.
According to Eq.~(\ref{def:Hpar}), this time is determined by the formula
\be
T(a_1,a_2) = A^{\vpsq}_0 \int_{a^{\vpsq}_1}^{a^{\vpsq}_2} da\,
\bigl[-{\mathcal U}_{\mbss{c}}(A^{\vpsq}_0 a)\bigr]^{-1/2}.
\label{def:t12}
\ee
The most important time intervals are
$T_{\mbsu{L}} = T(a_{\mbsu{t.L}},1)$
(from the left turning point to the present-day value of $a$),
$T_{\mbsu{C}} = T(1,a_{\mbsu{c.L}})$
(from $a=1$ to the left critical point),
$T_{\zeta} = T(a_{\mbsu{c.L}},a_{\mbsu{c.R}})$
(the interval between the critical points), and
$T_{\mbsu{R}} = T(a_{\mbsu{c.R}},a_{\mbsu{t.R}})$
(from the right critical point to the right turning point).
The values of $T_{\mbsu{L}}$, $T_{\mbsu{C}}$, and $T_{\mbsu{R}}$
can be calculated immediately by the formula (\ref{def:t12}).
The determination of the interval $T_{\zeta}$ requires an additional analysis
in view of the supposed smallness of the parameter $\zeta$.
To leading order in $\zeta$ one obtains
\be
T_{\zeta} = \frac{4}{3}\,t^{\vphu}_{\mbsu{g}}\left(
\frac{\sqrt{2|\tau_{\mbss{c}}|A_{\mbsu{c.L}}}}{|\tilde{w}^{\prime}(A_{\mbsu{c.L}})|} +
\frac{\sqrt{2|\tau_{\mbss{c}}|A_{\mbsu{c.R}}}}{|\tilde{w}^{\prime}(A_{\mbsu{c.R}})|}
\right)\zeta^{3/4},
\label{Tzappr}
\ee
where $t^{\vphu}_{\mbsu{g}}$ is the Compton time of the graviton [see Eq.~(\ref{def:tg})],
$A_{\mbsu{c.L(R)}} = A^{\vpsq}_0\,a_{\mbsu{c.L(R)}}$, and according to the above definitions
\be
\tilde{w}(A_{\mbsu{c.L}}) = \tilde{w}(A_{\mbsu{c.R}}) = 0\,.
\label{def:cpnt}
\ee

To calculate the absolute value of $T_{\zeta}$ one needs the value of the
unknown parameter $\zeta$.
The ratio of the graviton mass to the Planck mass seems to be a reasonable
candidate for this role because of its extreme smallness (see Table~\ref{tab:1})
and because both masses already enter the gravitational action of the TDR.
So, the quantity $\zeta^{(\mbsu{tent})} = m^{\vphu}_{\mbsu{g}}/m^{\vphu}_{\mbsu{P}}$
was taken as the tentative value of $\zeta$.
The values of the above-mentioned time intervals for the quasipotential of the TDR
with the parameter sets from Table~\ref{tab:1} are listed in Table~\ref{tab:2}.
The values of $T^{\,(\mbsu{tent})}_{\zeta}$ were calculated by the formula (\ref{Tzappr})
with $\zeta = \zeta^{(\mbsu{tent})}$.
As can be seen, these values of $T_{\zeta}$ are much less than the typical times of the
direct nuclear reactions ($10^{-22}$ s).
But it should be noted that the value of $\dot{a}^2$ rapidly increases inside
the interval between the critical points towards its center.
This dynamics is not taken into account in Eq.~(\ref{Tzappr}) derived for the
fixed interval $(a_{\mbsu{c.L}},a_{\mbsu{c.R}})$.

The absence of the zero-point singularity of the quasipotential and
the extremely small value of $T^{\,(\mbsu{tent})}_{\zeta}$ imply that
the Big Bang or the ``Big Bang plus Big Crunch'' scenarios of the
modern cosmology (see, e.g., Refs. \cite{Overduin04,Ryden17})
are replaced in the TDR by the cyclic ``small bang plus small crunch'' scenario.
Then $T_{\mbsu{C}}$ is the time remaining until the next small bang.
The quantity $T_{\mbsu{s.c.}} = 2 T_{\mbsu{L}} + T_{\mbsu{C}}$ is the time
that has passed since the last small crunch.
Notice that in spite of the word ``small'', both these events are rather dramatic
because they are related to the change, practically instantaneous,
of the matter density in more than 5 times.
The half-period $T_{1/2}$ of the oscillations of the scale factor
is the sum of all time intervals, i.e.
\be
T_{1/2} = T_{\mbsu{L}} + T_{\mbsu{C}} + T^{\vphu}_{\zeta} + T_{\mbsu{R}}\,,
\label{def:Thp}
\ee
but, as follows from Table~\ref{tab:2}, it is mainly determined by the value of
$T_{\mbsu{R}}$.

The last row of Table~\ref{tab:2} contains also the values of the ``age of the Universe''
$T_{\mbsu{U}}$ in the flat $\Lambda$CDM model which is defined as (see \cite{Peebles84})
\be
T_{\mbsu{U}} = T(0,1) = \frac{2}{3 H^{\vphu}_0 \sqrt{\Omega_{\Lambda}}}
\ln \Bigl( \bigl( 1 + \sqrt{\Omega_{\Lambda}} \bigr)/
\sqrt{\Omega_{\mbss{m}}}\Bigr)\,.
\label{TLCDM}
\ee
The values of $T_{\mbsu{U}}$ are close to each other for all data sets
and turn out to be quite close to $T_{\mbsu{L}}$,
however $T_{\mbsu{U}} > T_{\mbsu{L}}$ in all cases.
The calculated values of $T_{\mbsu{C}}$ and $T_{\mbsu{R}}$ have a large spread.
Probably the most important quantity $T_{\mbsu{C}}$ takes the values
from 17~Myr in the case of the binned data set WM57 from \cite{Farooq17}
to 1.8~Gyr in the case of the data set \cite{Yu18}.
The quantity $T_{\mbsu{R}}$ takes the values from 5.8~Tyr in the case of the
binned data set WM57 to 47.0~Tyr in the case of the full data set \cite{Farooq17}.
This spread is mainly explained by the fact that the $H(z)$ data used in the fit
of the quasipotential parameters
refer to the range of values of the scale factor beyond (or even far away from)
the region of integration in Eq.~(\ref{def:t12}) for these intervals.

\subsubsection{Properties of the dual matter
\label{sssec:pdm}}

\begin{figure}[]
\begin{center}
\includegraphics*[trim=2.5cm 15.5cm 0cm 2cm,clip=true,scale=0.5,angle=0]{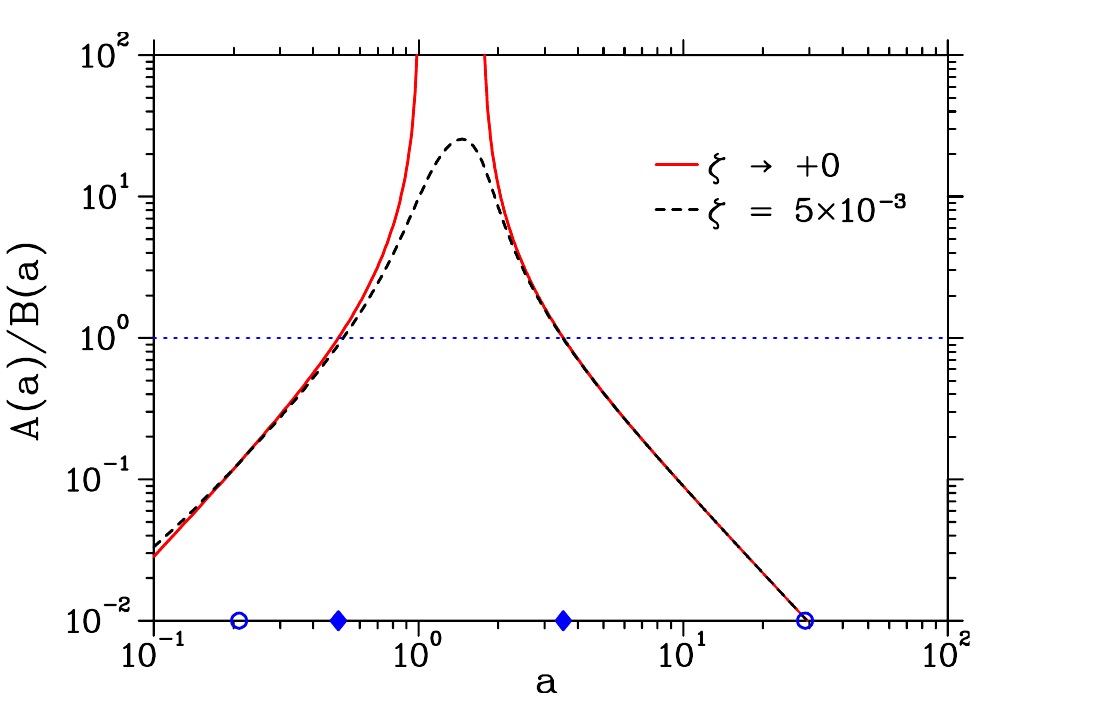}
\end{center}
\caption{\label{fig:ABwm57}
Dependence of the ratio of the scale factor $A$ to the lapse function $B$
on the reduced scale factor $a$.
The turning points of the quasipotential are indicated by the blue open circles.
The matching points where $A(a)/B(a)=1$ are indicated by the blue filled diamonds.
Calculation with parameters fitted to the binned data set WM57 from Ref.~\cite{Farooq17}.
}
\end{figure}

The action functional of the fields of the dual matter (\ref{actdm})
does not specify its properties. However, two important properties of this kind
of matter can be deduced from the general TDR equations.
First, as follows from Eq. (\ref{def:jmn}),
the energy-momentum tensor of the dual matter $\tilde{T}^{\vphu}_{\mu\nu}$
enters into the r.h.s. of the TDR equation of motion (\ref{eomsym}) with negative sign
[though it enters with the positive sign into the r.h.s. of
Eqs. (\ref{tempd}) and (\ref{eqsol1}) for the total energy-momentum tensor density].
It means that the interaction between the ordinary and the dual matter
has antigravitational (repulsive) character, but gravitational interaction
between the parts of the dual matter itself remains attractive.
One can also arrive at the same conclusions by considering the Newtonian limit
of the TDR equations.
The repulsive long-range interaction between the ordinary and the dual matter
can in principle serve as a mechanism of the formation of the cosmic voids
(see Refs. \cite{Weygaert11,Wilson23} and references therein).
But this issue requires a separate study.

The second property is related to the fact that the different metrics on one and the same
manifold can lead to the different speeds of light for those kinds of matter
which are governed by these metrics.
In the cosmological limit the metrics of the ordinary and dual matter
are determined by Eqs. (\ref{def:gmncl}) and (\ref{def:tgmncl}), respectively.
In terms of the Cartesian coordinates, the metric (\ref{def:gmncl}) gives
the following equation of the light cone for the ordinary matter
\be
 A^2 [(dx^1)^2 + (dx^2)^2 + (dx^3)^2] = B^2 (dx^4)^2.
\label{lconom}
\ee
The analogous equation for the dual matter is determined by the metric (\ref{def:tgmncl}).
It reads
\be
 B^2 [(dx^1)^2 + (dx^2)^2 + (dx^3)^2] = A^2 (dx^4)^2.
\label{lcondm}
\ee
From Eqs. (\ref{lconom}) and (\ref{lcondm}) it follows  that the light cone
of the ordinary matter lies inside the cone of the dual matter if $A^2/B^2 > 1$.
In this case, the superluminal motion of the particles of dual matter
with respect to the particles of ordinary matter becomes possible.
However, if the interaction between these kinds of the matter is reduced only
to the (anti)gravity [as follows from Eqs.~(\ref{def:actm}) with account of the remark
after Eqs.~(\ref{def:tttmn})], this superluminal motion is difficult to observe.
Notice that all this is not related to the theory of tachyons which was quite actively
developed and discussed in the past (see, e.g., Ref.~\cite{Recami86} and references therein).
Obviously, the situation becomes opposite if $A^2/B^2 < 1$.
In this case, the ordinary matter has the superluminal properties
with respect to the dual matter.
As can be seen from Table~\ref{tab:1}, the present-day values of $A^2/B^2$
for the different parametrizations of the quasipotential
are considerably greater than one.

The $a$-dependence of the ratio $A(a)/B(a)$ calculated with parameters fitted
to the binned data set WM57 from Ref.~\cite{Farooq17} is shown in Fig.~\ref{fig:ABwm57}.
The matching points where $A(a)=B(a)$
(in the given case, these are the points $a=0.50$ and $a=3.52$)
are indicated by the blue filled diamonds.
The inequality $A^2/B^2 > 1$ is fulfilled in the region between these points.
Outside this region, one has the inverse inequality $A^2/B^2 < 1$.
This picture is typical for all parametrizations listed in Table~\ref{tab:1}.

\section{Summary and conclusions}

In the paper a new version of the modified gravity theory is presented.
The theory is based on the vierbein formalism including two vierbein fields
which are connected with each other by the duality condition.
This condition contains the flat background metric $\gamma$, and thereby
reduces the number of the independent field variables of the theory
to the number of variables of the conventional GR.
Since the duality condition plays a crucial role in this theoretical scheme,
it gives the name to the whole approach: the theory of dual relativity (TDR).

The gravitational action functional in the TDR has a simple polynomial form.
As compared to the classical GR, it contains only two additional constants:
the graviton mass and the so-called mixing parameter $\zeta$.
Thus, the TDR is the theory with massive graviton.
The gravitational action of the classical GR is recovered
if the graviton mass and the parameter $\zeta$ are set equal to zero.
It is shown that the TDR provides the unambiguous definition
of the total energy-momentum tensor density of the closed physical system,
and that this tensor density satisfies the conservation law.

In the TDR it is assumed that there exist two kinds of matter:
the ordinary matter governed by the metric $g$ and the dual matter governed
by the metric $\tilde{g} = \gamma\,g^{-1} \gamma$.
Presence of the inverse metric $g^{-1}$ in the formula for $\tilde{g}$ implies,
and this is shown explicitly,
that the dual matter possesses antigravitational and (under certain conditions)
superluminal properties with respect to the ordinary matter.

The action functional of the TDR contains the term which formally corresponds to the
$\Lambda$-term in the GR. But this term and the analogous term for the dual matter
are included into the matter actions, and it is supposed that these terms
fully compensate contributions of the vacuum expectation values of the matter fields,
so that the effective value of the cosmological constant $\Lambda$ is exactly equal to zero.
This scheme formally solves the known problem related to this constant.

The cosmological limit of the TDR equations is considered.
In this limit, there are two classes of solutions of these equations:
with positive and with negative total energy density.
The solutions with positive energy density describe
the stable Universe (the Universe as a whole) governed by the averaged flat metric $\gamma$
if the certain conditions imposed on the parameters of the TDR equations are fulfilled.
It is shown that the existence of both ordinary and dual matter is necessary
for the fulfillment of these stability conditions.

The class of solutions with negative energy density describes
the domains of the Universe where the scale factor of the FRW metric oscillates.
It is supposed that our part of the Universe is one of such domains.
For this class of solutions,
the model formula for the dependence of the Hubble parameter $H$
on the redshift $z$ has been obtained. The parameters of this formula are found
from the fit to the available $H(z)$ data. It is obtained that the TDR gives
a better description of the $H(z)$ data as compared to the flat $\Lambda$CDM model.
The important consequences of the obtained results are, first, their incompatibility
with the standard Big Bang model and, second, the existence of two critical points
determining a sharp change of the scale factor in the course of its oscillations.
These results can be interpreted in terms of the
cyclic scenario ``small bang plus small crunch''.

Among the open issues of the theory one can emphasize the following.
Though the concept of stable Universe seems to be very attractive,
and the TDR equations formally admit the respective solutions,
the possibility of the existence of stable or quasistable
domain structure is not obvious.
The questions remain regarding the real existence of the dual matter,
its quantitative contribution in the description of the data,
and the possible connection of this kind of matter with the formation of cosmic voids.
In spite of the recovery of the gravitational action of the classical GR
at the zero values of the graviton mass and the mixing parameter in the TDR,
there are the known problems in the theories with massive graviton
\cite{Hinterbichler,vanDam70,Zakharov70,Vainshtein72,Babichev10,Boulware72}
which were not considered here.
Finally, as was noted in Sec.~\ref{sec:basdef},
a more consistent approach implies that the two-vierbein scheme used in the TDR
should be deduced from the gauge theory based on the extended Lorentz symmetry
considered in Ref.~\cite{T08}.
These and other issues imply a need for further investigations.


\bibliographystyle{apsrev4-1}
\end{document}